\documentclass[fleqn,usenatbib]{mnras}

\usepackage{newtxtext,newtxmath}

\usepackage[T1]{fontenc}

\DeclareRobustCommand{\VAN}[3]{#2}
\let\VANthebibliography\thebibliography
\def\thebibliography{\DeclareRobustCommand{\VAN}[3]{##3}\VANthebibliography}

\usepackage[dvipsnames]{xcolor}
\usepackage{newtxtext,newtxmath}
\usepackage[T1]{fontenc}
\usepackage{ae,aecompl}
\usepackage{adjustbox}
\usepackage{xcolor}
\usepackage{graphicx}	
\usepackage{amsmath}	

\title[Period changes of Cepheids]{Study of changes in the pulsation period of 148 Galactic Cepheid variables}

\author[G. Cs\"ornyei et al.]{
G. Cs\"ornyei,$^{1,2,3}$\thanks{E-mail: csornyei.geza@csfk.org}
L. Szabados,$^{1,4}$
L. Moln\'ar,$^{1,4,5}$
B. Cseh,$^{1}$
N. Egei,$^{1}$
Cs. Kalup,$^{1,6}$
V. Kecskem\'ethy,$^{1,6}$
\newauthor
R. K\"onyves-T\'oth,$^{1,7}$
K. S\'arneczky,$^{1}$
and R. Szak\'ats$^{1}$
\\
$^{1}$Konkoly Observatory, ELKH Research Centre for Astronomy and Earth Sciences, Konkoly Thege Mikl\'os \'ut 15-17, 1121 Budapest, Hungary\\
$^{2}$Max-Planck-Institut f\"ur Astrophysik, Karl-Schwarzschild-Str. 1, 85741 Garching, Germany\\
$^{3}$Physics Department, Technische Universit\"at M\"unchen, James-Franck-Str. 1., 85741 Garching, Germany \\
$^{4}$CSFK Lend\"ulet Near-Field Cosmology Research Group, Konkoly Thege Mikl\'os \'ut 15-17, 1121 Budapest, Hungary\\
$^{5}$ELTE E\"otv\"os Lor\'and University, Institute of Physics, P\'azm\'any P\'eter s\'et\'any 1/A, 1117 Budapest, Hungary\\
$^{6}$ELTE E\"otv\"os Lor\'and University, Department of Astronomy, P\'azm\'any P\'eter s\'et\'any 1/A, 1117 Budapest, Hungary\\
$^{7}$Department of Optics \& Quantum Electronics, University of Szeged, D\'om t\'er 9, 6720 Szeged, Hungary
}

\date{Accepted XXX. Received YYY; in original form ZZZ}

\pubyear{2021}

\begin{document}
\label{firstpage}
\pagerange{\pageref{firstpage}--\pageref{lastpage}}
\maketitle

\begin{abstract}
Investigating period changes of classical Cepheids through the framework of O-C diagrams provides a unique insight to the evolution and nature of these variable stars. In this work, the new or extended $O-C$ diagrams for 148 Galactic classical Cepheids are presented. By correlating the calculated period change rates with the {\em Gaia} EDR3 colours, we obtain observational indications for the non-negligible dependence of the period change rate on the horizontal position within the instability strip. We find period fluctuations in 59 Cepheids with a confidence level of 99\%, which are distributed uniformly over the inspected period range. Correlating the fluctuation amplitude with the pulsation period yields a clear dependence, similar to the one valid for longer period pulsating variable stars. The non-negligible amount of Cepheids showing changes in their $O-C$ diagrams that are not or not only of evolutionary origin points toward the need for further studies for the complete understanding of these effects. One such peculiar behaviour is the large amplitude period fluctuation in short period Cepheids, which occurs in a significant fraction of the investigated stars. The period dependence of the fluctuation strength and its minimum at the bump Cepheid region suggests a stability enhancing mechanism for this period range, which agrees with current pulsation models.
\end{abstract}




\section{Introduction}
\label{Sec:1}
The classical Cepheid variable stars (hereafter Cepheids) are objects of fundamental importance for both stellar astrophysics and extragalactic distance determination. These variables represent a brief phase in the post-main-sequence evolution of intermediate-mass (4-12 M$_{\odot}$) central helium burning stars, which populate the classical instability strip in the Hertzsprung--Russell (H--R) diagram. Their importance lies in the periodic variability of these objects: the characteristic feature of their pulsation, the period, allows for the calibration of the cosmic distance ladder and an independent estimation of the Hubble constant via the famous period-luminosity (PL) relation \citep{Freedman2001, Riess2019} or its reddening free formulation (Wesenheit function). Measuring the exact value of the period and the rate of its change can also give implications on the physical parameters of these stars and their evolution using pulsation \citep{Bono1999, Marconi2010} and evolutionary models that take into account various recently discovered effects, e.g. realistic core overshooting and meridional mixing from rapid rotation \citep{Maeder2001, Anderson2014}. It has been shown that, apart from the pulsation period, which is obviously the most crucial property for these variables, light curve parameters (amplitude, skewness and acuteness) can play an important role in the estimation of physical properties \citep{Bellinger2020}, however, the role of period changes on the precision of these estimations has not been investigated. Since Cepheids in different locations within the instability strip exhibit different period change rates, these values can be used to infer the probable value of physical properties of the stars, as well \citep{Turner2006}.

The pulsation period of Cepheids can change for multiple reasons. In most cases, the physical process behind these changes is stellar evolution, i.e. the movement of the Cepheid on the H--R diagram. This trajectory through the instability strip is associated with changes in the physical properties, which in turn change the period of pulsation: if the star evolves towards the cool edge of the instability strip, the period increases, while it decreases when evolving towards the hot edge. Superimposed on the curve corresponding to stellar  evolution, low-frequency quasi-cyclic period variations can also appear, e.g., in the case of S~Vul \citep{MahmoudSzabados1980}, several long-period Cepheids in the LMC \citep{Rodriguez-Segovia2021} or for many Cepheids mentioned in this article. These variations are due to period and phase fluctuations accumulating in time and building up a phase-lag that will appear as a change in the pulsation period \citep{Detre1965}. The timescale of these fluctuations varies within a wide range, from decades to even centuries. Due to these variations, period determination using $O-C$ diagrams can become uncertain, thus other methods might be necessary for this purpose when the baseline of the $O-C$ diagrams is not long enough to reliably separate the signals originating from the evolution and the fluctuations \citep{LombardKoen1993}. In most of the cases, long term photometry allows for the reliable separation of the stellar evolution from the fluctuations on the $O-C$ diagram. However, in some cases, as for BY~Cas and DX~Gem, the time span covered by observations and the timescale of fluctuations are comparable, which can even lead to the misidentification of the Cepheid's crossing number, as it was shown by \cite{Berdnikov2019c}. Although now there are photometric data from long enough timescales for the effects of evolution and fluctuations to be separated reliably for bright Cepheids, fluctuations can still prove a problem for the period change determination of fainter Cepheids that are observed less regularly. Physical processes behind such fluctuations could be either short-term variations in the atmospheric structure \citep{DeasyWayman1985}, or minor changes in helium abundance gradients in upper layers of Cepheids \citep{Cox1998}. The century long temporal coverage of light curves not only allows the investigation of period changes through the simple methodology of $O-C$ diagrams \citep{Sterken2005}, but it can give us an unprecedented insight into the nature of these fluctuations as well, which facilitates the refinement of current pulsation models. 

Apart from the evolutionary changes and the random fluctuations of the pulsation period, binarity of the Cepheid can also cause observable effects in the $O-C$ diagram. One of these is the light-time effect (LiTE), which arises from the orbital motion of the variable star. This effect can be easily distinguished from the evolutionary effect, since it results in a periodic feature in the $O-C$ graph. The orbital modulation can be detected in the $O-C$ diagram of binary Cepheids when the orbital period is sufficiently long, because in the case of short orbital periods the amplitude of this cyclic pattern would be too small to recognize. A very clear indication that a superimposed wavelike signal is caused by the LiTE instead of fluctuations is the stability of the signal; since the fluctuation waves are caused by accumulating random processes, the shape of the resulting quasi-cyclic signal will get distorted, however, as the LiTE is caused by a deterministic process, its shape will remain stable over time. The importance of light time effect lies in the fact that it can be used for a rough estimation of the orbital parameters. So far only a few Cepheids have been found to show LiTE (AW Per and RX Aur, \citealt{Szabados1992} and \citealt{Vinko1993}), with a few more awaiting spectroscopic confirmation.

Another effect characteristic of binary Cepheids is a phase jump or phase slip seen in the $O-C$ diagram. A phase jump in the $O-C$ diagram has a stepwise, while the phase slip has a sawtooth-like shape; in the first case the period remains the same while the phase of the pulsation suddenly changes, while in the second case, the period of pulsation changes rapidly, then returns to the previous value after a given amount of time (period jump-rejump). Such phase jumps have been found in the $O-C$ diagrams of several Cepheids (e.g. \citealt{Szabados1989}). One of the most prominent examples of Cepheids exhibiting such features in their $O-C$ diagrams is Polaris \citep{Turner2005}, which displayed a parabolic $O-C$ trend corresponding to a continuously increasing period that was interrupted by a phase jump after which the period continued to increase with the previous rate, with a significant phase difference. 

The idea that phase jumps appear only in the $O-C$ diagrams of binary Cepheids is based solely on empirical data. Each Cepheid showing such an effect is either a member of a known spectroscopic binary or suspected to have a companion based on independent evidence. Moreover, no phase jump was ever detected for any well-studied single Cepheid. The size of a phase jump is usually in the order of several hundredths of the pulsation period. Unfortunately, there is no theoretical explanation for the occurrence of such phase jumps or slips. For Polaris, \cite{Turner2005} suggested the brief interruption of the Cepheid's regular evolution by a short-lived blueward evolution associated with a small change in the average radius of the star, or alternatively, by a sudden increase in the mass of the Cepheid. However, these effects appear to be too spontaneous and peculiar to explain the phase jumps generally, especially when the phase jump occurs multiple times for a given star, which was observed to occur among Cepheids. \cite{Szabados1992} suspects that the occurrence of phase jumps is governed by the orbital motion. In this case, a possible explanation for the phase jumps could be the perturbing effect of the companion star that exerts increased influence on the upper layers of the atmosphere of the Cepheid (where the pulsation takes place) near periastron passage. The stepwise structure of the $O-C$ diagrams in the case of phase jumps contradicts current evolutionary models, which predict continuous change of pulsation period otherwise.

Even though only a few Cepheid variables were observed in the programs of the \emph{Coriolis}, \emph{MOST}, \emph{Kepler} and \emph{CoRoT} space missions, the advent of high precision space photometry brought improvements in the period change studies of these variables as well. Based on the observations of the Solar Mass Ejection Imager (SMEI) instrument onboard the \emph{Coriolis} satellite, \cite{Spreckley2008} connected slow changes in the light curve of the Polaris to a quite fast evolutionary phase and by supplementing the existing $O-C$ diagram suggested that another phase jump occurred. \cite{Berdnikov2010} studied the period changes of short period Cepheids based on SMEI photometry and detected random fluctuations in four of the observed stars. By analysing the \emph{MOST} observations of SZ Tau and RT Aur, \cite{Evans2015} revealed cycle-to-cycle variations that are a function of the pulsation phase. The stability of their pulsation was also investigated and, besides finding temporal variations on a scale of decades, it was pointed out that the pulsation of overtone Cepheids is more erratic than that of fundamental mode variables. In the case of the \emph{Kepler} mission, there was only one Cepheid in the observed field, V1154 Cyg. \cite{Derekas2012, Derekas2017} provided a detailed analysis of the data and found cycle-to-cycle fluctuations on the $O-C$ diagram, which were attributed to instabilities in the light-curve shape caused by convection and hot spots based on the suggestion of \cite{Neilson2014}. \cite{Poretti2015} performed the fluctuation tests similar to the work by \cite{Evans2015} on the \emph{CoRoT} data for seven Cepheids, and found small cycle-to-cycle variations. Since these space missions covered only a few Cepheids, our understanding on these small fluctuations is limited, which is expected to change through future works based on the data from the currently ongoing \emph{TESS} project \citep{Plachyetal2021}. 

In this work, our primary aim is to expand the collection of existing $O-C$ diagrams of known binary Cepheids by using century-long photometric observational data of these variables. To get a more complete set of diagrams, the initial set of variables was extended with stars that, to our current knowledge, do not belong to binary systems.

The structure of this paper is as follows. In Section~\ref{sec:two} we describe the available data and the differences of our method from the standard $O-C$ method. In Section~\ref{sec:three} we discuss the results obtained for Cepheids that exhibited some peculiarities on their $O-C$ diagrams individually. Then, in Section~\ref{sec:4} we analyse our sample collectively, first by placing the period change rates into broader context by comparing the results with previous works, then by discussing the period fluctuations in general. Finally, in Section~\ref{sec:four} we summarize our results and conclude the paper.

\section{Data and the method of the analysis} 
\label{sec:two}
\subsection{Data and observations}
To achieve the near century long coverage of the temporal behaviour of the pulsation period, we attempted to acquire all available processed photometric data for every star. The first half of the 20th century was covered by the Harvard College Observatory Plate database (DASCH project, \citealt{DASCH}, for stars on the northern galactic hemisphere) and by the observations of the American Association of Variable Star Observers (AAVSO, \citealt{aavso}, for suitably bright Cepheids). The later part of the century was covered by various articles from the literature, and by the earlier works of \cite{Szabados1989}, \cite{Szabados1991} and \cite{Berdnikov1997}. The complete set of references for articles used for the individual Cepheids can be found in Table~\ref{tab:allceps}. To complete this dataset with more recent observations, we used the photometric data obtained by the \textit{Hipparcos} \citep{hipp}, the All Sky Automated Survey (ASAS, \citealt{ASAS}), INTEGRAL-OMC \citep{IOMC}, KELT\footnote{\url{https://exoplanetarchive.ipac.caltech.edu/}}, ASAS-SN \citep{asassn1, asassn2}, the Optical Gravitational Lensing Experiment (OGLE, \citealt{OGLE}), the Solar Mass Ejection Imager (SMEI, \citealt{SMEI}), as well as the Kamogata Sky Survey \citep{kws} projects. To supplement these data, we have monitored some Cepheids with the 60/90\,cm Schmidt telescope at Piszk\'estet\H{o} Observatory. The photometric data acquired through this telescope can be found in Table \ref{tab:piszkes}.

\begin{table}
\begin{center}
\begin{tabular}{ccc}
Cepheid & JD & V \\
\hline
YZ Aur & 2458806.402 & 9.8815 \\
YZ Aur & 2458807.618 & 9.8637 \\
YZ Aur & 2458823.473 & 10.392 \\
YZ Aur & 2458824.418 & 10.129 \\
YZ Aur & 2458828.623 & 10.052 \\
YZ Aur & 2458855.319 & 10.683 \\
... & ...  & ... \\
\end{tabular}
\end{center}
\caption{List of observations taken at the Piszk\'estet\H{o} Observatory. The brightness values are differential magnitudes except for X~Lac, for which we used previously taken unpublished measurements which are not transformed to the international system. The full list of observations is available in the online supplement of the article.}
\label{tab:piszkes}
\end{table}

\subsection{Construction of the {\boldmath$O-C$} diagrams}

We followed the changes in the pulsation period of individual Cepheids using the method of $O-C$ diagrams \citep{Sterken2005}. To analyse the photometric datasets of each Cepheid, we applied Discrete Fourier Transformation (DFT, \citealt{Deeming1975}) on the measurements, for which we used the Period04 software \citep{Lenz2005}. We analysed the data from the various sources separately, to account for differences in circumstances valid for individual observational series, by keeping the frequencies and the relative phases of the harmonics fixed to a pre-calculated parameter set, which were based on the dataset with the widest coverage and most data points, while allowing the amplitudes to change during the fitting procedure. 

As a first step of the analysis, every set of observations were split into smaller subsets. For each survey, depending on the temporal coverage of the data, 300-450 day long temporal bins were defined depending on the length of the pulsation period, in which each data point was moved to a new subset. The folded light curves of each previously created subset of measurements were calculated, which then were used to determine the observed ($O$) epoch values of the average (zero point) brightness on the brightening edge of the light curve. To determine the $O$ values, we fitted the phase folded light curves corresponding to the different epochs with the previously calculated Fourier components representative of the data source at hand, by only allowing the mean brightness and phase offset of the model light curve to change. The $O$ values were then simply calculated from the phase offset of the best fit model. The use of the median brightness on the ascending branch was motivated based on the Fourier fit qualities: for longer and longer pulsation periods, the complexity of the light curve increases, which in turn means that more and more Fourier components will need to be taken into account. For older, less accurate datasets the fitting of the high order components could become uncertain, which in turn could introduce a bias to the inferred $O-C$ values, when they are based on the epoch of maximum brightness. However, it was found that the rising branch itself, and the epoch of average brightness on it can be fitted with suitable accuracy even in such cases, thus providing a viable alternative for datasets with lower amount of observations. The advantage of using the epoch of average brightness over that of the maximum becomes even more important for Cepheids with pulsation periods between 8-10 days, due to the secondary maximum present on the light curve. It was also found by \cite{Derekas2012}, that the most accurate $O-C$ values can be obtained through measuring the epoch of average brightness on the rising branch. The average brightness value was determined as the baseline for the sinusoidal components at the Fourier fitting of the separate datasets. By this definition the brightness value at the zero point is slightly different for various datasets as it will account for the differences between the different surveys and measurement systems, thus the resulting $O-C$ points will be comparable. Although the phase folding method inevitably decreases the resolution of the resulting $O-C$ curve, the precision of the results increases, since the error of the phase calculation will decrease significantly.

\begin{figure*}
\centering
\includegraphics[scale=0.55]{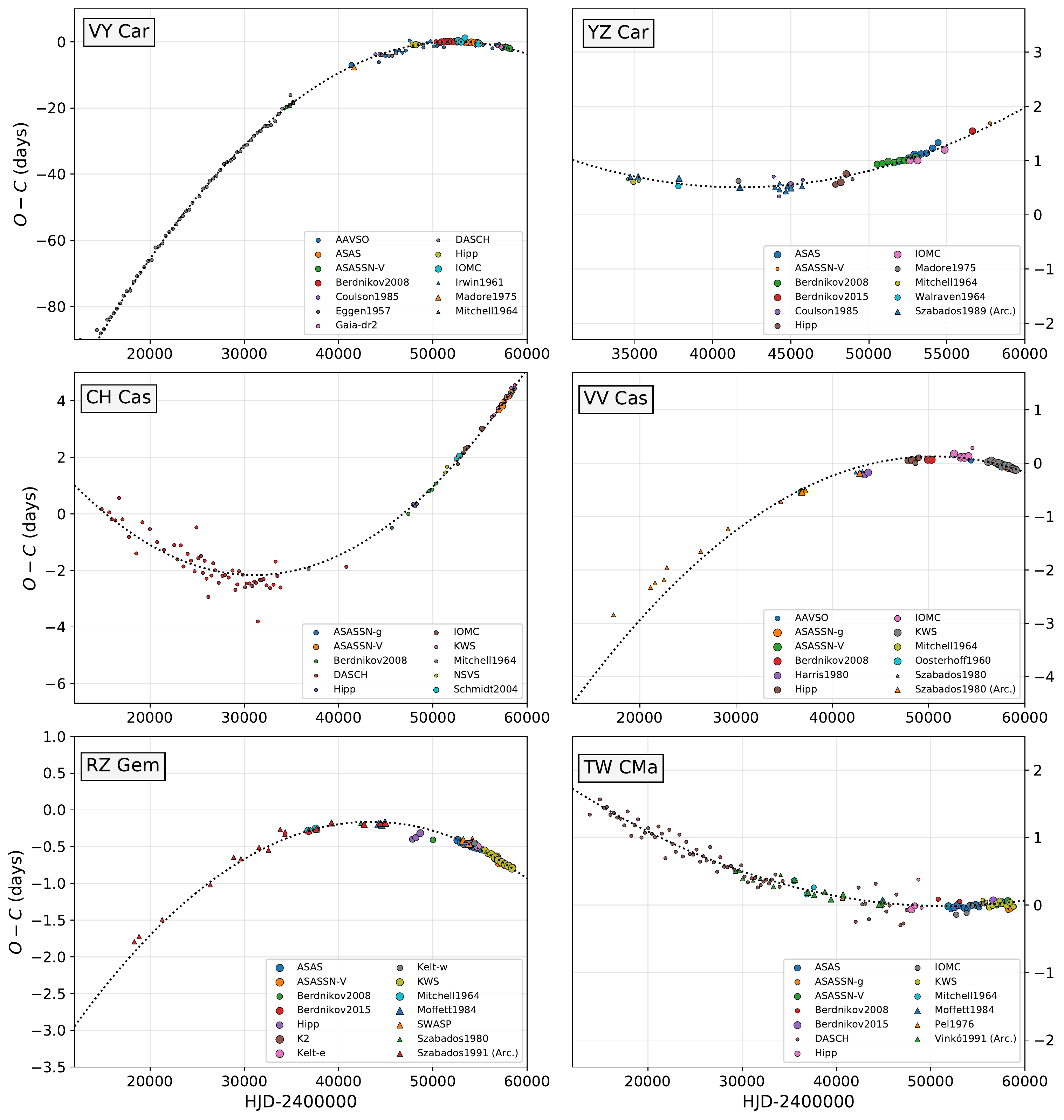}
\caption{Sample Cepheids from our set showing no features apart from the evolutionary signal. The size of the datapoints indicate the inverse of the uncertainty with a non-linear scaling.}
\label{fig:sample}
\end{figure*}

To complete the construction of the $O-C$ diagram, the errors of the individual $O-C$ differences had to be calculated. The most important factor that governs the precision of the individual $O-C$ values is the number of photometric data points used for their derivation. Since we only fit the average brightness and the phase shift of a given light curve shape, even a low number of data points is enough for the $O-C$ calculation, however to make sure that the results we obtain are meaningful, we set the lower limit on this number to 8. Another factor that needs to be taken into account when estimating the accuracy of the $O-C$ points is the phase distribution of photometric data points in each of the folded light curves. Obviously, if the phase curves created with the method explained above were well-covered, i.e. the calculated phases covered the possible range in an approximately uniform manner, then the standard least-squares method could be used to fit the known light-curve shape to the phase curve, which would allow for the straightforward calculation of the covariance matrix. However, if the phase-curves were not well covered, which occurred frequently in the case of long period variables observed in a series of short-cadence observation runs (e.g. with IOMC), then the above method could significantly underestimate the error of the $O-C$ differences. To avoid this problem, the errors have been calculated by applying bootstrapping throughout the fitting procedure; at each step 75\% of the data points have been randomly selected, and fitted with the precomputed light-curve shape. This step was carried out a hundred times for each phase curve. The final uncertainty of the individual $O-C$ residuals was calculated as the square root of the sum of overall phase shift and light-curve zero point variance.

The $O-C$ derivation method inherently contains a minor bias through the use of Fourier decomposition for the fitting, as it can obviously just approximate the actual shape of the light curve, which could cause biases in cases when a dozen or more Fourier components are required for the description. Although these biases can be minimized properly by measuring the moment of the average brightness on the rising branch instead of that of the maximum, a more precise and general method could be achieved by using a non-parametric light curve fitting instead of the Fourier decomposition. Nevertheless, since we used the same amount of Fourier components for every dataset of the individual stars, these biases affect each of the reductions similarly, thus they did not lead to noticeable artefacts between the different data sources.

\section{Remarks on individual Cepheids}
\label{sec:three}

The $O-C$ curves corresponding to evolutionary changes are usually represented as parabolic trends, which correspond to constant rate of period change (for a sample of $O-C$ diagrams showing solely the effects of stellar evolution, see Fig.\ref{fig:sample}). There have been numerous studies from the last decades showing such behaviour (e.g. \citealt{Parenago1956}, \citealt{Szabados1983}, \citealt{Berdnikov1997}). The rate of period change depends on the position of the Cepheid within the instability strip and the crossing number, and is usually modelled by a linear function of time, which rates are well approximated by current pulsation and evolutionary models \citep{Anderson2016}.

However, there is no physical model that restricts the period change of a star to be linear, since the luminosity and the temperature themselves change in a nonlinear manner. Hence, as it has been argued by \cite{Fernie1990}, in some cases higher order polynomial fits are necessary for certain $O-C$ diagrams, which indicate not only the change of period itself, but variations in the rate of period change as well. Such behaviour is common among Cepheids, significant deviations can be observed from parabolic fits to the $O-C$ plot in numerous cases. Determining the period change rate for a large set of Cepheids allows for the statistical study of this phenomenon, as well as giving way for the refinement of existing period-based empirical relations. However, there are quite a few classical Cepheid variables that exhibit peculiar $O-C$ diagrams, for which the statistical description is not yet available due to the small number of such objects, thus these stars have to be treated separately.

In this section we present and discuss the $O-C$ diagrams exhibiting features which could not be explained by evolutionary changes with superimposed relatively small amplitude fluctuations, such as phase jumps or slips, or large amplitude waves that partially, if not completely, obscure the effect of evolution on the diagram. In the large sample of analysed Cepheids, we found 16 of such $O-C$ diagrams, which are presented below.

\subsection{Cepheids showing wavelike $O-C$ structure}
Out of the 16 Cepheids that exhibit peculiar $O-C$ diagrams, we found 9 stars that show a remarkable wavelike modulation. Such wavelike signals can arise either due to the presence of a companion star through LiTE, or due to the fluctuations present in the pulsation of the Cepheid, in which case the modulation can even obscure the effect of evolution on the $O-C$ diagram. To model these wavelike signals we simultaneously fitted the underlying parabolic trend corresponding to stellar evolution and the superimposed modulation (except for UY~Mon, where we could not calculate such a fit reliably, see Sec.~\ref{sec:UYMON}). By this procedure, we fitted a single sinusoidal waveform for each Cepheid, except for IR~Cep, where involvement of one additional sinusoidal component is motivated based on the systematic deviations present in the residuals calculated with respect to the parabolic plus single sinusoidal model fit. In the case of the other Cepheids, we either did not find any additional signals in the residual data after fitting the parabola and the modulation, or due to the randomness and possible fluctuation origin of the modulation (as in the case of VZ~CMa and DX~Gem, see Sec.~\ref{sec:VZCMa} and \ref{sec:DXGem}) we decided that adding further components would not change the outcome of the analysis significantly.

For these Cepheids, similarly to the other Cepheids showing evolutionary changes, we then statistically analysed the residuals calculated by various models (for the constant period, the linear period-change, and the linear period change plus sinusoidal signal cases) using F-test, in order to determine the significance of the wavelike modulation in the modelling. The final F-statistics and the associated p-values can be found in the summary table in the Appendix (Table~\ref{tab:ftest}). In a later section (Sec.~\ref{sec:fluct}) we then used these solutions along with those of Cepheids showing evolutionary changes for the collective analysis.

\subsubsection{RX~Aurigae}
\label{sec:RXAur}
RX~Aur is a bright intermediate period Cepheid, whose binarity was long suspected based on the slope of its colour-colour loop and the LiTE-like nature of its $O-C$ diagram \citep{Szabados1988}. The total mass of the system inferred based on this diagram did not rule out the presence of a companion \citep{Szabados1988}, and it is further supported by the RV measurements of \citet{Gorynya1996}, who marked the star as a possible binary.

\begin{figure} 
\centering
\includegraphics[width = \linewidth]{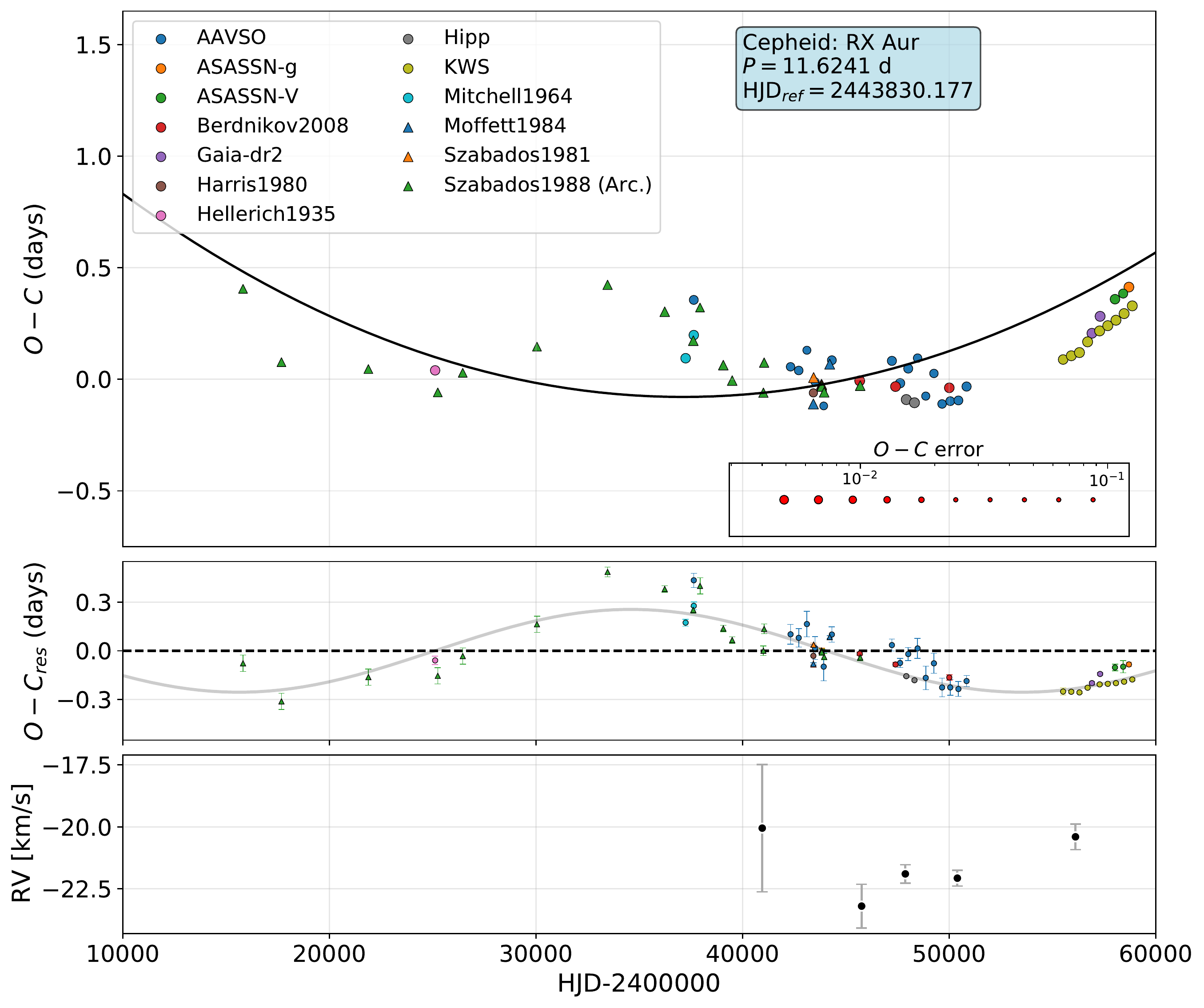}
\caption{$O-C$ diagram of RX Aurigae, exhibiting a parabolic change corresponding to evolution of the Cepheid and a wavelike signature superimposed (top panel), the residual $O-C$ values with the corresponding fit (middle panel) and the available mean RV values.}
\label{fig:RXAur1}
\end{figure}

By extending the $O-C$ diagram with more recent photometric measurements we confirm the LiTE explanation of the $O-C$ diagram (Fig.~\ref{fig:RXAur1}). The $O-C$ diagram was calculated using the following elements:
$$ C_{\textrm{med}} = 2443830.177 + 11.6241 \cdot E $$

\begin{figure} 
\centering
\includegraphics[width = \linewidth]{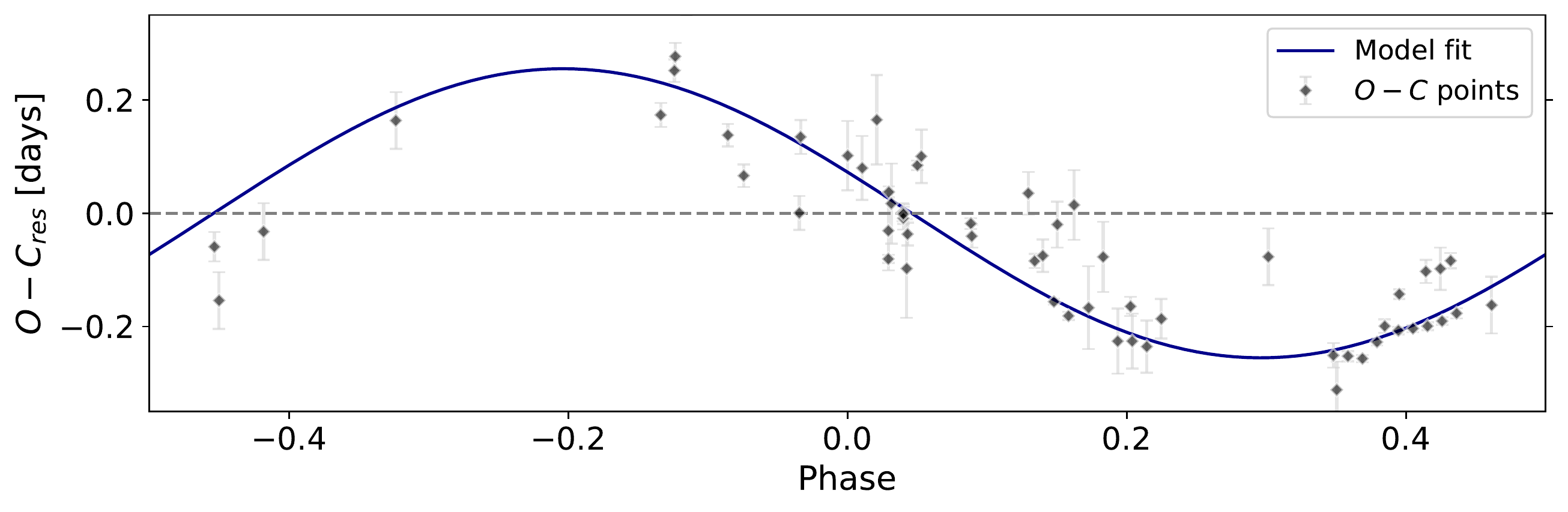}
\caption{Phase diagram of the residual $O-C$ variation of RX Aur. The elements used for the calculation of the phase diagram were $T_0 = 2442312.825$ and $P = 37928.243$ d.}
\label{fig:RXAur2}
\end{figure}

Fitting the evolutionary trend and the superimposed modulation (Fig.~\ref{fig:RXAur2}) yielded an orbital period of $P_{\rm orb} = 37928 \pm 7678$ d $= 103.84 \pm 21.02$ yr and an amplitude of $A = 0.255 \pm 0.107$ d, which are approximately 1.92  and 1.15 times larger, respectively, than the previously calculated values \citep{Szabados1988}. Since, as seen in Fig.~\ref{fig:RXAur1}, the maximum of the variation in the residual terms is only covered once by our dataset, a more detailed and reliable orbital parameter determination is not yet possible, as the few data points near that extremum do not provide sufficient constraint for the fitting procedure. However, the knowledge of the orbital period and semi-major axis (amplitude) allows for the determination of the total mass of the system through the
\begin{equation}
\label{eq:Kepler}
\left(\frac{A}{\sin i} \right)^3 P_{\textrm{orb}}^{-2} = M_1 + M_2 
\end{equation}
formula (since $A = a \sin i$, with $a$ being the semi-major axis of the orbit), where $A$ is measured in astronomical units, $P_{\textrm{orb}}$ in years. The resulting parameters are as follows:
\begin{equation*}
\begin{split}
 a \sin i = (44.152 \pm 18.515) \textrm{ AU }\\
 = (6.605 \pm 2.769) \cdot 10^9 \textrm{ km }\\
 (M_1 + M_2 ) \sin^3 i = (7.999 \pm 6.239) M_{\odot}. 
\end{split}
\end{equation*}

The resulting mass is consistent with the values inferred by \citet{Kervella2019a} with the assumption that the orbital plane is oriented along or near the line of sight ($i \gtrsim 60^{\circ}$), and the obtained orbital period also falls into the range proposed by them. However, no significant signal can be extracted from the available RV data (from the measurements of \citealt{Schmidt1974}, \citealt{Barnes1987}, \citealt{Gorynya1996}, \citealt{Imbert1999} and \citealt{Borgniet2019}) displayed in Fig.~\ref{fig:RXAur1}, which show a low scatter ($\sigma_{\textrm{RV}} \sim 1.155$ km/s) and their distribution is not aligned in phase with the suspected LiTE either, hence it is not possible to reach a solid conclusion on the origin of the observed $O-C$ signal yet.

\subsubsection{VZ Canis Maioris}
\label{sec:VZCMa}
VZ~CMa is a short period Cepheid whose pulsation mode was investigated in various studies, yielding different results: \citet{Kienzle1999} classified this star as an overtone pulsator, while according to \citet{Groenewegen2000} this Cepheid is a fundamental mode pulsator. Its multiplicity was also studied in various articles: it was first suspected by \citet{Stobie1979} and \citet{Szabados1993} based on its color indices and amplitude ratios respectively, then was finally confirmed spectroscopically by \citet{Szabados1996}. The $O-C$ diagram of the object was studied by \cite{Berdnikov1994}, who found it to be parabolic.

\begin{figure} 
\centering
\includegraphics[width = \linewidth]{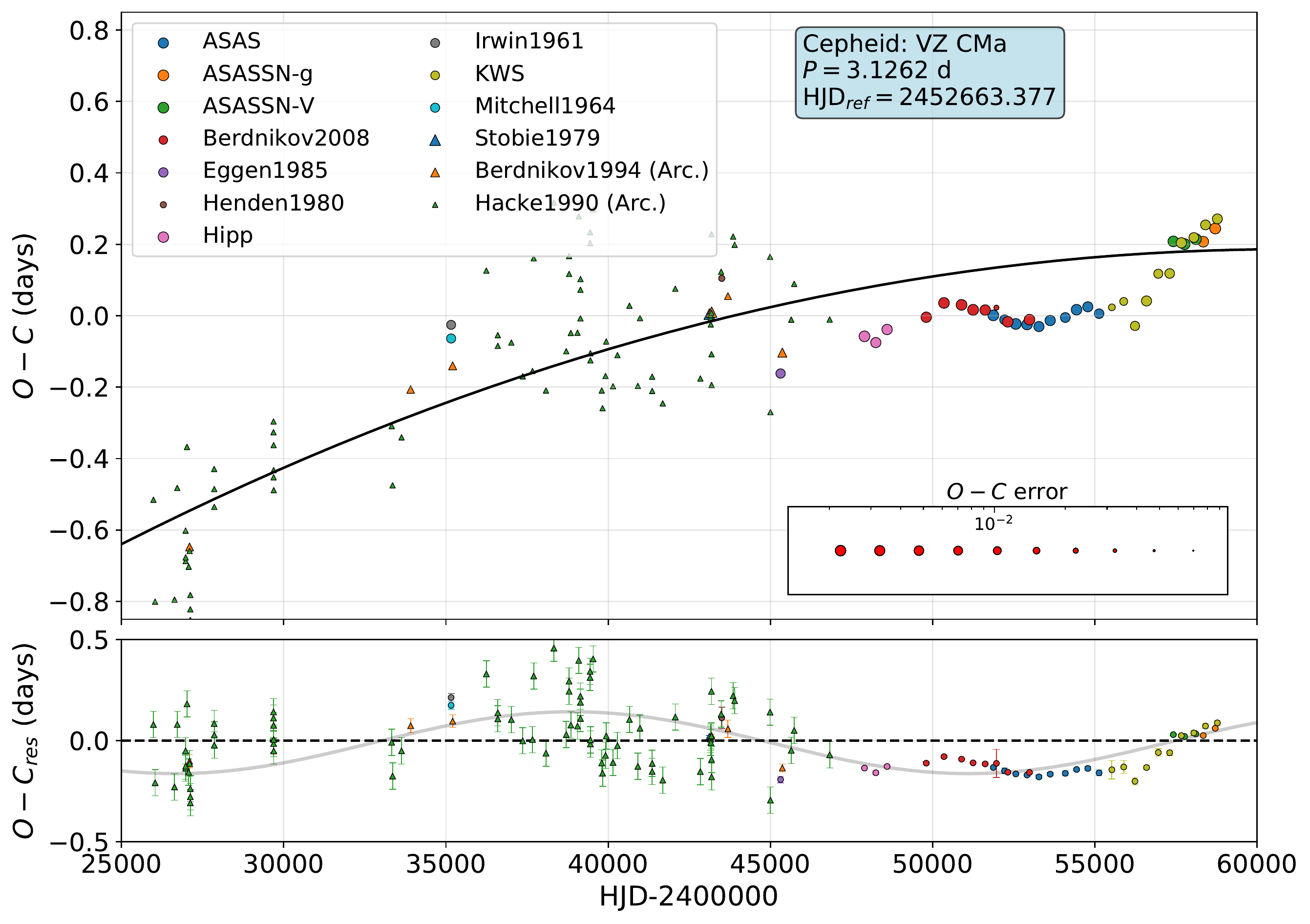}
\caption{$O-C$ diagram of VZ~CMa exhibiting a larger amplitude quasi-periodic signal, which could either be attributed to LiTE (which however, would not explain the systematic deviations of data points at HJD 2450000 on the lower panel) or to period fluctuations.}
\label{fig:VZCMa}
\end{figure}

By extending the previously available data with the archival measurements of \cite{Hacke1990}, and with that of modern surveys, we found that the $O-C$ diagram of VZ~CMa exhibits a large amplitude wave (Fig.~\ref{fig:VZCMa}), which was interpreted as a parabolic pattern before, due to the lack of sufficient data. The $O-C$ diagram was calculated with the following elements:
$$ C_{\textrm{med}} = 2452663.377 + 3.126151 \cdot E $$

Fitting the parabolic trend and the wavelike signal simultaneously, we estimated the period of the modulation to be 24475 d (67.01 yr) and its peak-to-peak amplitude to be approximately 0.3 d. Although the values inferred here could indicate a LiTE origin for the signal, the unusual, non-sinusoidal shape points towards the presence of fluctuations in the pulsation. This ambiguity is not lifted by the RV datasets either, as the combined observations of \citet{Stobie1979} and \citet{Kienzle1999} cover a much shorter timespan than the period of the signal, henceforth we cannot conclude whether this wave originates from binarity or period fluctuations.

\subsubsection{BY~Cassiopeiae}
\label{sec:BYCas}
BY~Cas is short period Cepheid with a long history of cluster membership and binarity studies. Membership in NGC~663 was first suggested by \citet{Malik1965}, which was contradicted by the later study of \citet{Usenko2015} citing large color excess, parallax and RV differences, concluding that BY~Cas is a foreground star, in agreement with the result of \citet{Anderson2013} too. However, according to the latest analysis of \citet{Glushkova2015}, this Cepheid shows connection not only to its nearest cluster NGC~663, but also to the clusters NGC~654, NGC~659, and the association Cas~8, too. BY~Cas was also revealed to have a companion with an orbital period of $\sim$560 d \citep{Gorynya1994}.

\begin{figure} 
\centering
\includegraphics[width = \linewidth]{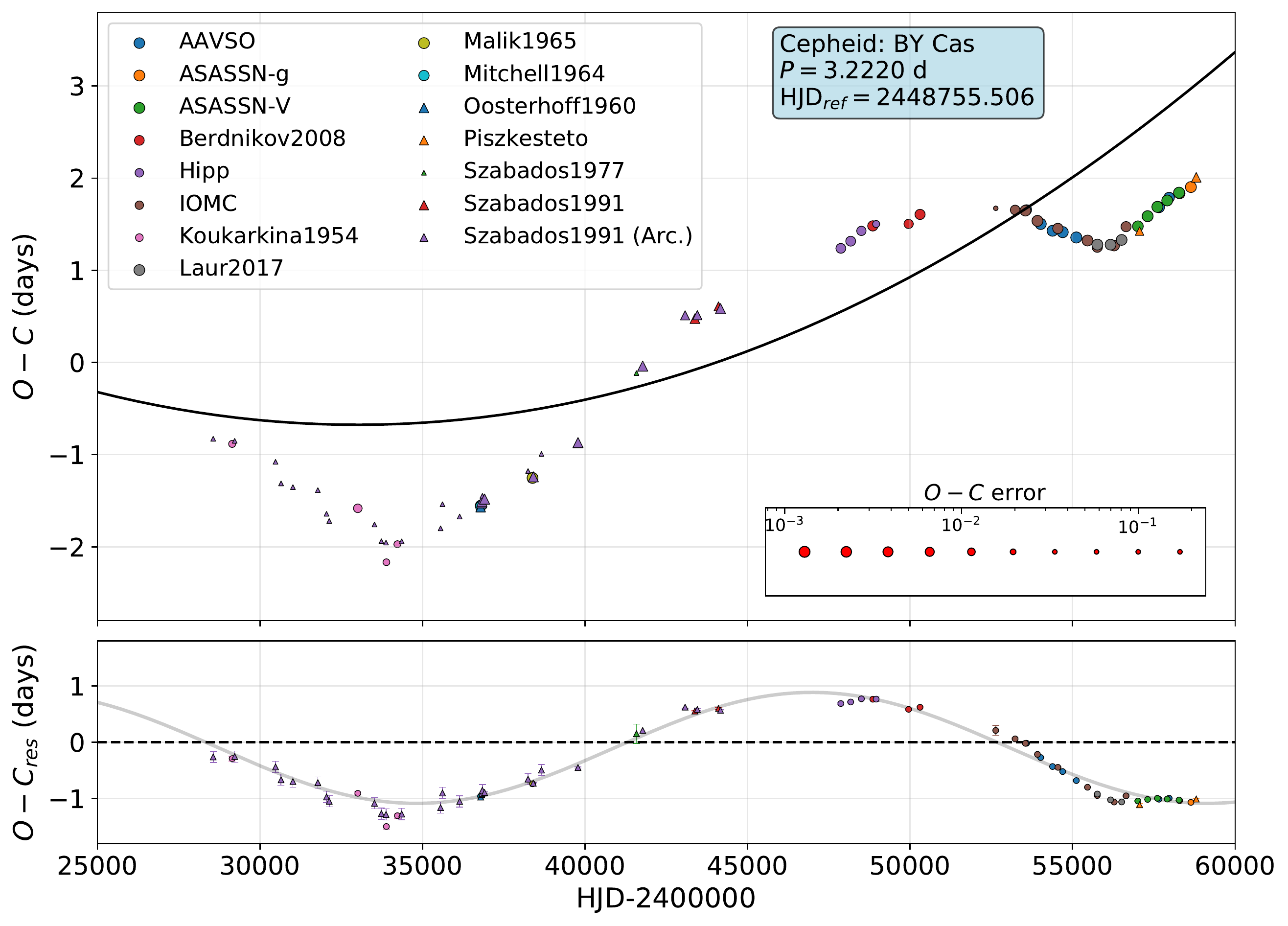}
\caption{$O-C$ diagram of BY~Cas, showing an evolutionary change with a large amplitude wavelike signal superimposed.}
\label{fig:BYCas}
\end{figure}

The period changes of this variable were investigated by \cite{Szabados1991} and \cite{Berdnikov1994}, and more recently by \cite{Berdnikov2019a}. The former analysis described the variations as a sequence of sudden period changes, while the latter work proposed evolutionary changes with a wavelike pattern superimposed. Our independent analysis showed that the $O-C$ diagram can indeed be explained by evolutionary changes with a wavelike signal on top of it (Fig.~\ref{fig:BYCas}). The presented $O-C$ diagram was calculated with the following elements:

$$ C_{\textrm{med}} = 2448755.506 + 3.222019 \cdot E $$

Using the parabolic fit we estimated the period change rate to be 0.338~s$\cdot$yr$^{-1}$, which is two orders of magnitude smaller than the one estimated by \cite{Berdnikov1994}, and places this object in the group of third crossing Cepheids, in agreement with the findings of \cite{Berdnikov2019a}. The best explanation for the superimposed modulation is the period fluctuation, which is relatively common for these Cepheids (as shown in the later sections).

\subsubsection{IR~Cephei}
IR~Cep is a short period Cepheid which was first studied in detail by \cite{Szabados1977}, revealing strong period change. This Cepheid was found to exhibit a peculiar light curve: according to \cite{KunSzabados1988}, the light curve of this Cepheid rather resembles that of larger amplitude Cepheids. Thus it was suggested that IR~Cep might be a classical Cepheid, but with a bright companion whose constant contribution to the emitted light reduces the observable amplitude of the variable, however, no close binary component was found by the RV analysis of \cite{Marschall1993}. \cite{KunSzabados1988} also suggested the possible association of this Cepheid to Cep~OB2.

\begin{figure} 
\centering
\includegraphics[width = \linewidth]{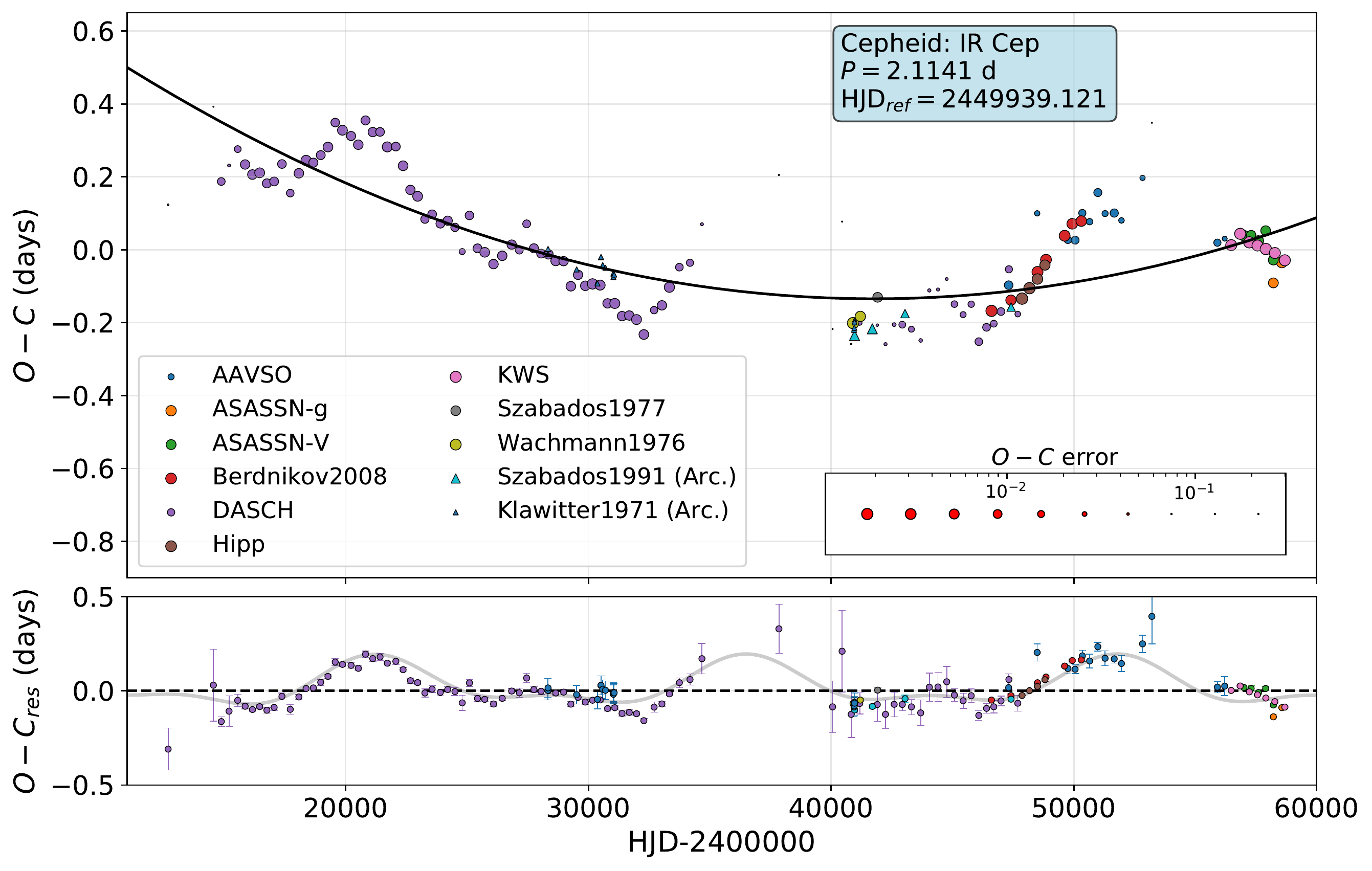}
\caption{$O-C$ diagram of IR~Cep (top panel). In the bottom panel the parabolic term corresponding to the evolution of the Cepheid had been subtracted. A simple Fourier fit of two harmonics can be seen plotted on the residual data points. The current amount of data points did not allow the addition of more harmonics terms, but this fit illustrates well that the shape of the residual signal remains relatively unchanged during the time interval covered by observations.}
\label{fig:IRCep}
\end{figure}

IR~Cep turned out to be peculiar in its pulsation as well: this Cepheid was classified as a first overtone pulsator by \cite{Groenewegen2000}, but \cite{Szabo2007} showed that a first overtone classification would lead to excessive phase lag discrepancy for this star. They note however, that according to its position on the phase lag diagram it would quite naturally fit in the class of fundamental mode Cepheids.

Based on the acquired photometric measurements the corresponding $O-C$ diagram has been drawn, which covers more than a century (Fig.~\ref{fig:IRCep}). The presented diagram was calculated using the following elements:
$$ C_{\textrm{med}} = 2453453.824 + 2.114088 \cdot E $$

The rate of period change is 0.042~s$\cdot$yr$^{-1}$, while the approximate period and peak-to-peak amplitude of the seemingly periodic signal are 15150 d (41.47 yr) and 0.29 d. The origin of this wavelike signal is uncertain, as the amplitude could allow for LiTE, which is also favoured by the more stable form of the signal, however due to the short temporal coverage of the RV data obtained by \cite{Marschall1993} and \cite{Gorynya1998} this could not be confirmed spectroscopically, hence the fluctuation origin cannot be excluded either.

\subsubsection{V532~Cygni}
V532~Cyg is a short period Cepheid known to be a binary system member. The presence of a companion star was first suspected by \cite{Madore1977} based on the shape of its colour-colour loop, then a similar conclusion was reached by \cite{Usenko1990} and \cite{Szabados1991} later on. This was later confirmed by the RV observations of \cite{Gorynya1996}, finding a 388~d orbital period.

\begin{figure} 
\centering
\includegraphics[width = \linewidth]{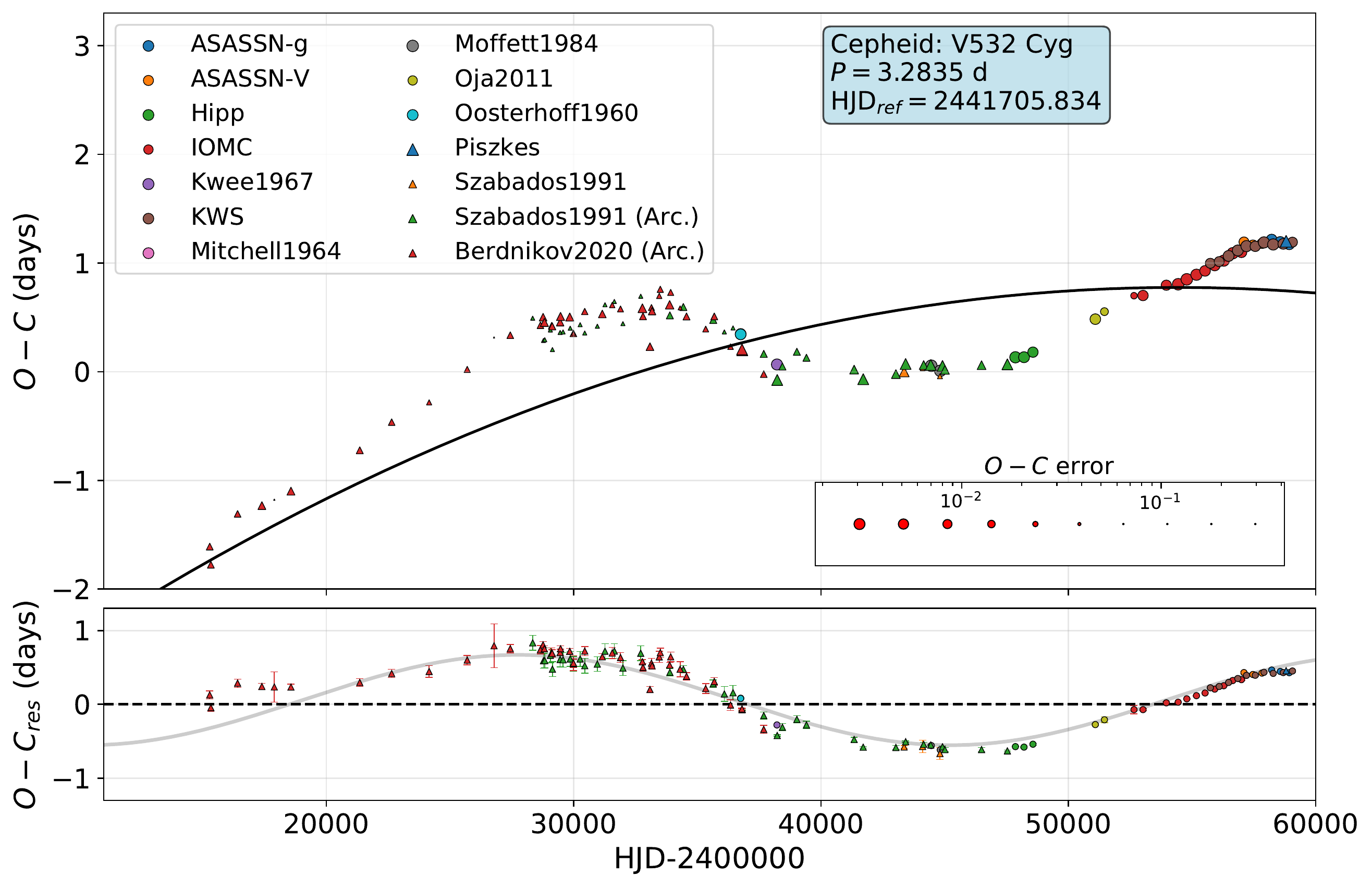}
\caption{$O-C$ diagram of V532~Cyg exhibiting an evolutionary trend with large amplitude wavelike modulation superimposed.}
\label{fig:V532Cyg}
\end{figure}

The period changes of this Cepheid were studied several times in the literature: \cite{Szabados1977, Szabados1991} first described the $O-C$ diagram as a sequence of period jumps, however, it was later shown that it can be described by evolutionary changes and a large wavelike signal superimposed \citep{Berdnikov2019b}. The latter behaviour of the $O-C$ diagram is also confirmed by our independent study, as shown in Fig.~\ref{fig:V532Cyg}. The presented $O-C$ diagram was calculated using the following elements:
$$ C_{\textrm{med}} = 2441705.834 + 3.283536 \cdot E $$

As it is shown in the $O-C$ diagram, the period of the Cepheid is indeed decreasing due to its evolution with a rate of $-$0.104~s$\cdot$yr$^{-1}$, which places it among the second crossing Cepheids. The large amplitude ($A = 0.612 \pm 0.013$ d) of the superimposed signal suggests that the fluctuation of the period is a suitable explanation.

\subsubsection{DX~Geminorum}
\label{sec:DXGem}
DX~Gem is a short period Cepheid which was for long considered one of the few first crossing Cepheids due to its large period change rate. The first period analysis on this Cepheid was done by \cite{Szabados1977,Szabados1991}, in which a series of phase jumps that appeared to be in accord with the binarity \citep{Burki1985} of the object were suggested as interpretation. However, \cite{Berdnikov2019c} showed that the $O-C$ diagram can be described by an evolutionary change with a wavelike signal superimposed.

\begin{figure} 
\centering
\includegraphics[width = \linewidth]{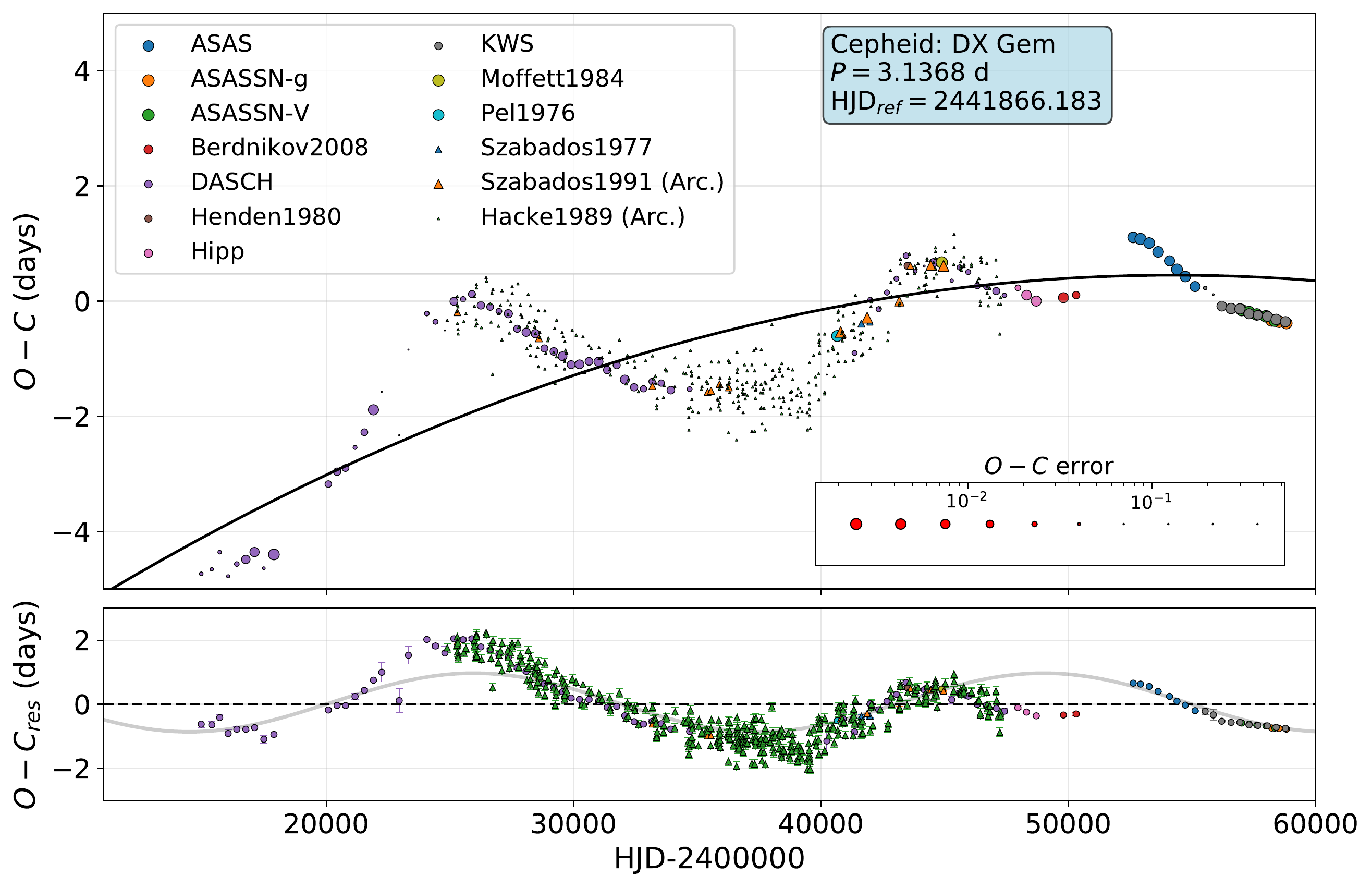}
\caption{$O-C$ diagram of DX~Gem (top panel). In the bottom panel the parabolic term corresponding to the evolution of the Cepheid had been subtracted. The remaining residual term exhibits a wavelike pattern, which is unstable and too large in amplitude to be connected to LiTE.}
\label{fig:DXGem}
\end{figure}

According to our independent analysis, the $O-C$ diagram of DX~Gem indeed cannot be explained by evolutionary changes alone (see Fig.~\ref{fig:DXGem}). Our $O-C$ diagram was calculated using the following elements:
$$ C_{\textrm{med}} = 2441866.183 + 3.136779 \cdot E $$

From the obtained fit it is apparent that the Cepheid exhibits a shortening period with a rate of $-$0.186~s$\cdot$yr$^{-1}$, hence it belongs to the class of second crossing Cepheids. The residual large amplitude modulation ($A = 0.917 \pm 0.027$ d) cannot be explained by LiTE, thus the most reasonable explanation for it is the presence of fluctuations in the period.

\subsubsection{UY~Monocerotis}
\label{sec:UYMON}
UY~Mon is a short period Cepheid that was rather neglected in the past: it was first correctly classified by \cite{Imbert1981}, as it was considered as an eclipsing binary before, and it has not been studied in terms of RV data and cluster membership yet. The period changes of this Cepheid were investigated before by \cite{Berdnikov1997}, who found an increasing period.

In our analysis, we extended the $O-C$ diagram compiled by \cite{Berdnikov1997} in both directions using the Harvard Observatory plate archives and more recent observations. The $O-C$ diagram was calculated using the following elements:
$$ C_{\textrm{med}} = 2432500.253 + 2.398185 \cdot E $$

According to our results, the $O-C$ diagram of UY~Mon looks like a slanted sinusoidal signal, with a possible period longer than 45000d ($\sim$120~yr). As the amplitude of this oscillation is small compared to the length of the period, the possibility of LiTE cannot be excluded. However, the time interval covered by the compiled data does not allow us to determine whether this signal is indeed periodic, thus further photometric and long-term RV observations are necessary for a firm conclusion.

\begin{figure} 
\centering
\includegraphics[width = \linewidth]{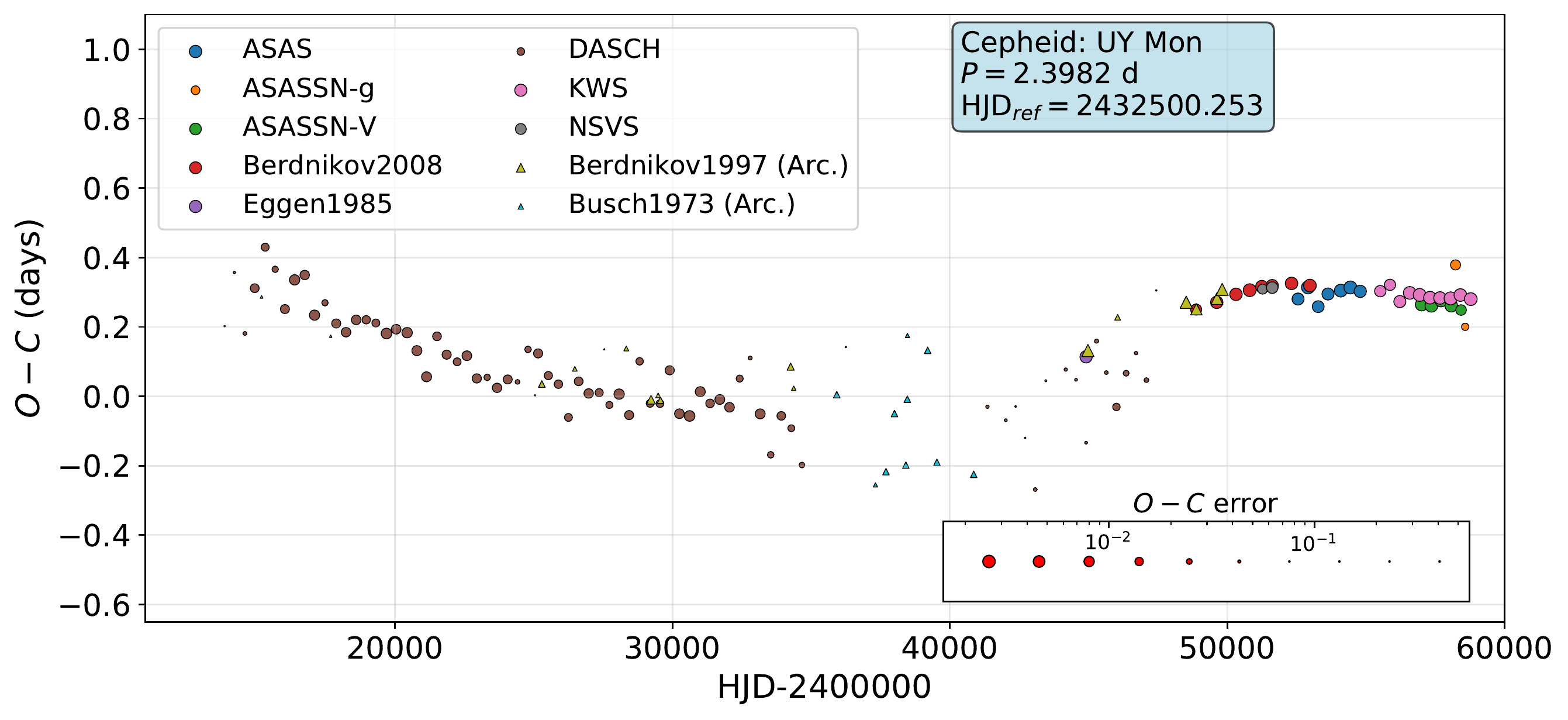}
\caption{$O-C$ diagram of UY~Mon exhibiting a long-term low amplitude wavelike signal, which could be a result of LiTE caused by a formerly unknown binary component in the vicinity of the Cepheid.}
\label{fig:UYMon}
\end{figure}

\subsubsection{Y~Ophiuchi}
Y~Oph is an intermediate period Cepheid that was frequently analysed in the past. It is a known binary with an orbital period of $\sim$1222.5 d \citep{Szabados1989}: the presence of the companion was first suspected by \cite{Pel1978} based on photometry, as Y~Oph appeared too blue for its pulsation period. However, no such companion was detected from the IUE spectra \citep{Evans1992b} nor through NACO lucky imaging \cite{Gallenne2014}. On this basis, its true nature remains elusive.

\begin{figure} 
\centering
\includegraphics[width = \linewidth]{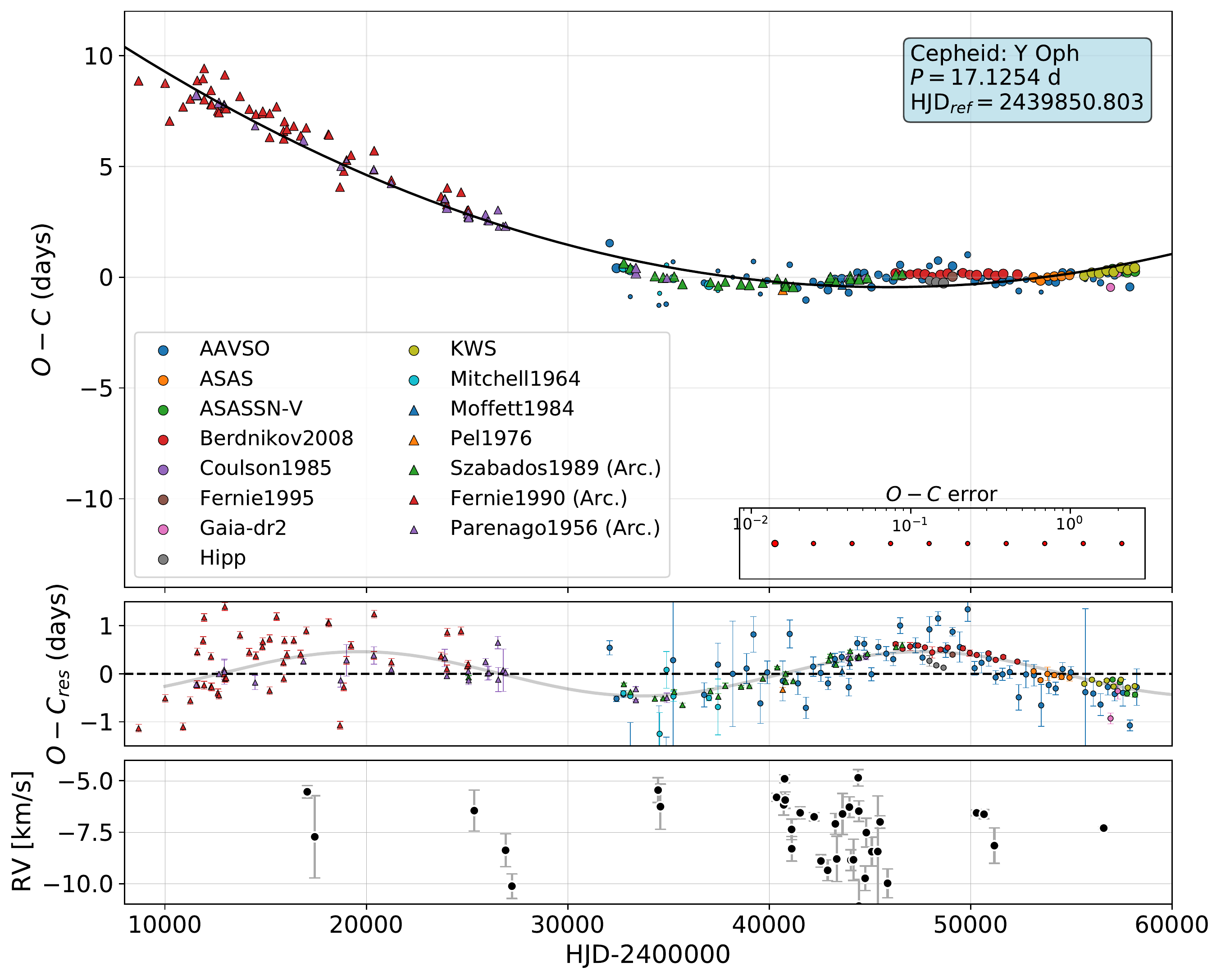}
\caption{$O-C$ diagram of Y~Oph exhibiting a wavelike trend corresponding to the LiTE superimposed on the secular period change of the Cepheid (top panel), the residual $O-C$ diagram obtained after subtracting the parabola (middle panel) and the available RV data points whitened for both the pulsation and the orbital motion of the Cepheid (bottom panel).}
\label{fig:YOph}
\end{figure}

By collecting not only the more recent measurements, but the results from previous works on this Cepheid as well \citep{Parenago1956, Szabados1989, Fernie1990} we compiled an $O-C$ diagram spanning almost a century (Fig.~\ref{fig:YOph}). To calculate the $O-C$ diagram, we assumed the following elements:
$$ C_{\textrm{med}} = 2439848.803 + 17.127827 \cdot E. $$

\begin{figure} 
\centering
\includegraphics[width = \linewidth]{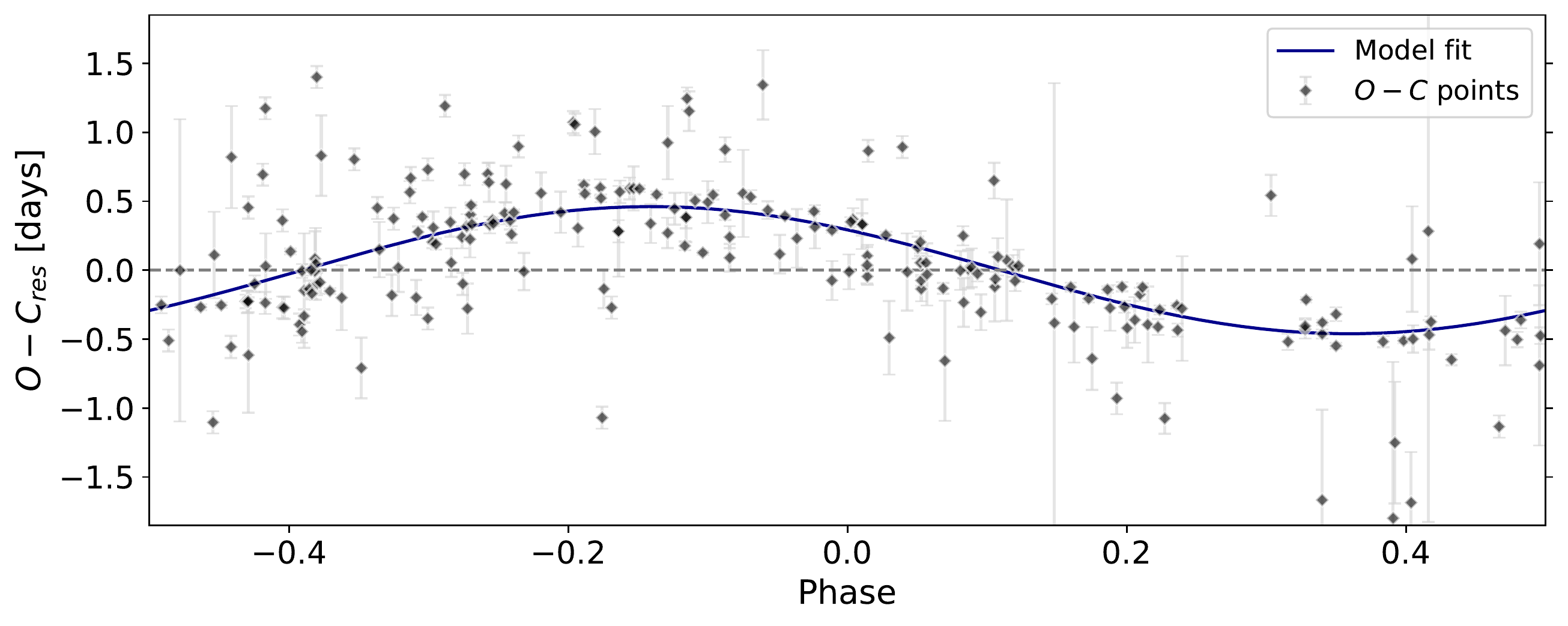}
\caption{Phase diagram of the residual $O-C$ data points of Y~Oph and the obtained fit. The elements used for calculating the phase diagram were $T_0 = 2451548.800$ and $P = 27948.469$ days.}
\label{fig:YOph_ph}
\end{figure}

The obtained $O-C$ diagram can be described by a combination of the parabolic trend corresponding to the evolution of the Cepheid and a wavelike signal superimposed, with the period of the modulation being $P=27948 \pm 1164$ d $=76.52 \pm 3.19$ yr and the amplitude $A=0.461 \pm 0.024$ d (Fig.~\ref{fig:YOph_ph}). If we assume that this periodic variation originated from LiTE, then the obtained amplitude corresponds to a semi-major axis of $a \sin i = 79.82 \pm  4.19$ AU $=(1.194 \pm 0.063)\cdot 10^{10}$ km, which agrees well with the long period of this wave. Although the duration of the covered time interval does not allow for a detailed fit of the residual curve, the accuracy of the simple sinusoidal fit suggests that the assumed orbit has a low inclination, which also agrees well with this period--semi-major axis pair. Since the period is much longer than the one obtained for the previously suspected companion, the light-time interpretation of the wavelike signal would suggest that a third component is present in the system of Y~Oph, but no signal with such a period could be extracted from the presently available RV data (bottom panel of Fig.~\ref{fig:YOph}). Henceforth, we cannot yet confirm the presence of the third component, however, as we will show it in Sec.~\ref{sec:4}, the amplitude of the modulation we found is larger and can be considered an outlier compared to the fluctuation amplitudes of similar period Cepheids, which also favours the new companion and the LiTE explanation.

\subsubsection{AW~Persei}
AW~Per is a short period Cepheid in a binary system. The binary nature of the star was discovered by \cite{MillerPreston1964} based on Ca II and K line measurements, with the orbit being investigated later on by \cite{WelchEvans1989}, \cite{Vinko1993}, and then by \cite{Evans2000}. This object was also found to exhibit LiTE that could be matched to RV observations \citep{Szabados1991}.

\begin{figure} 
\centering
\includegraphics[width = \linewidth]{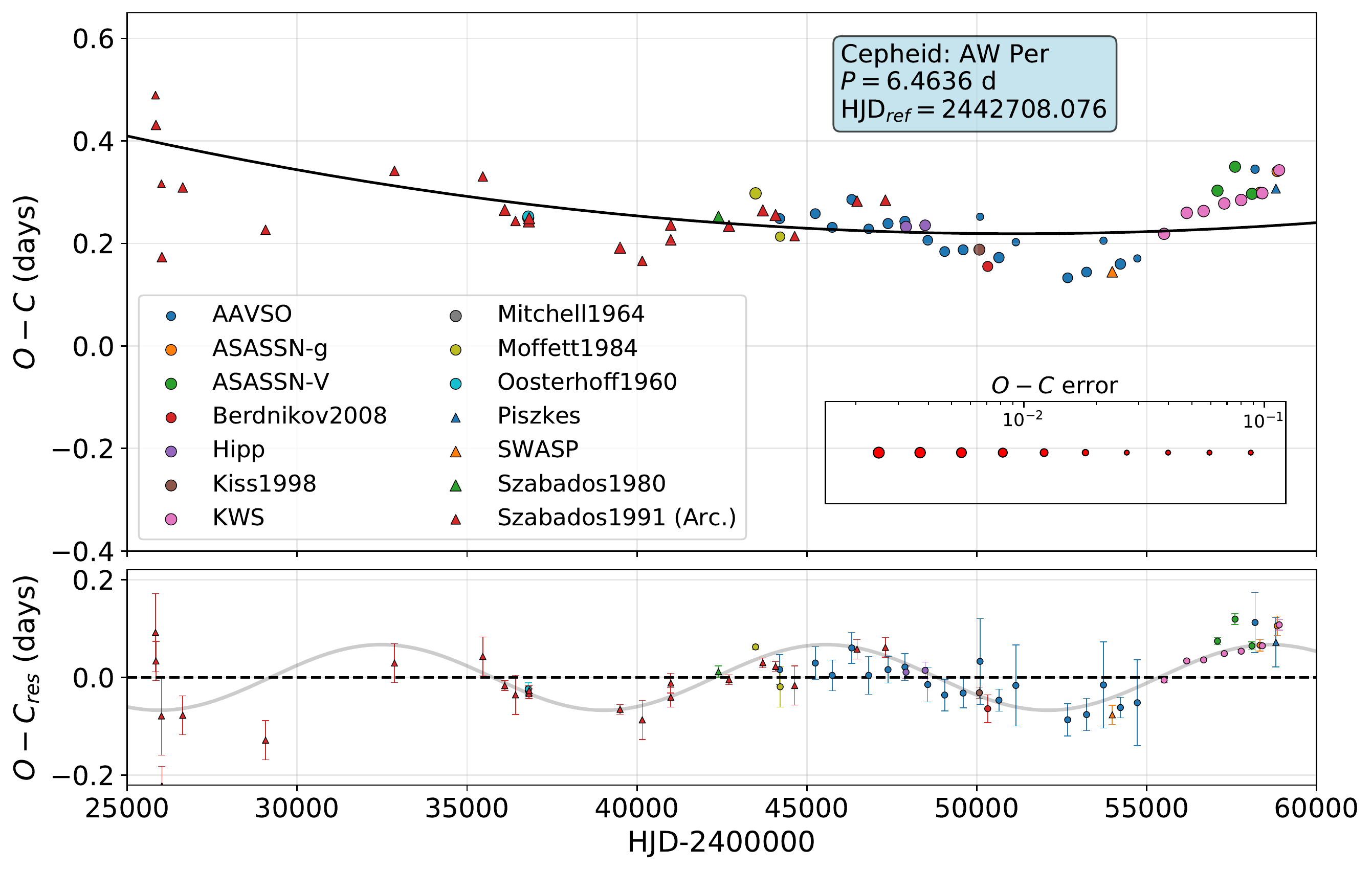}
\caption{$O-C$ diagram of AW~Per exhibiting a wavelike trend corresponding to the LiTE superimposed on the secular period change of the Cepheid (top panel) and the residual $O-C$ diagram obtained after subtracting the parabola (bottom panel). The fitting of this wavelike trend can be used for the determination of orbital elements as described in the text.}
\label{fig:AWPer1}
\end{figure}

To further investigate the signal created by the LiTE we extended the $O-C$ dataset of \cite{Szabados1991} with more modern observations from various sources. The resulting $O-C$ diagram (Fig.~\ref{fig:AWPer1}) was calculated with the following elements:
$$ C_{\textrm{med}} = 2442708.076 + 6.463688 \cdot E .$$

\begin{figure} 
\centering
\includegraphics[width = \linewidth]{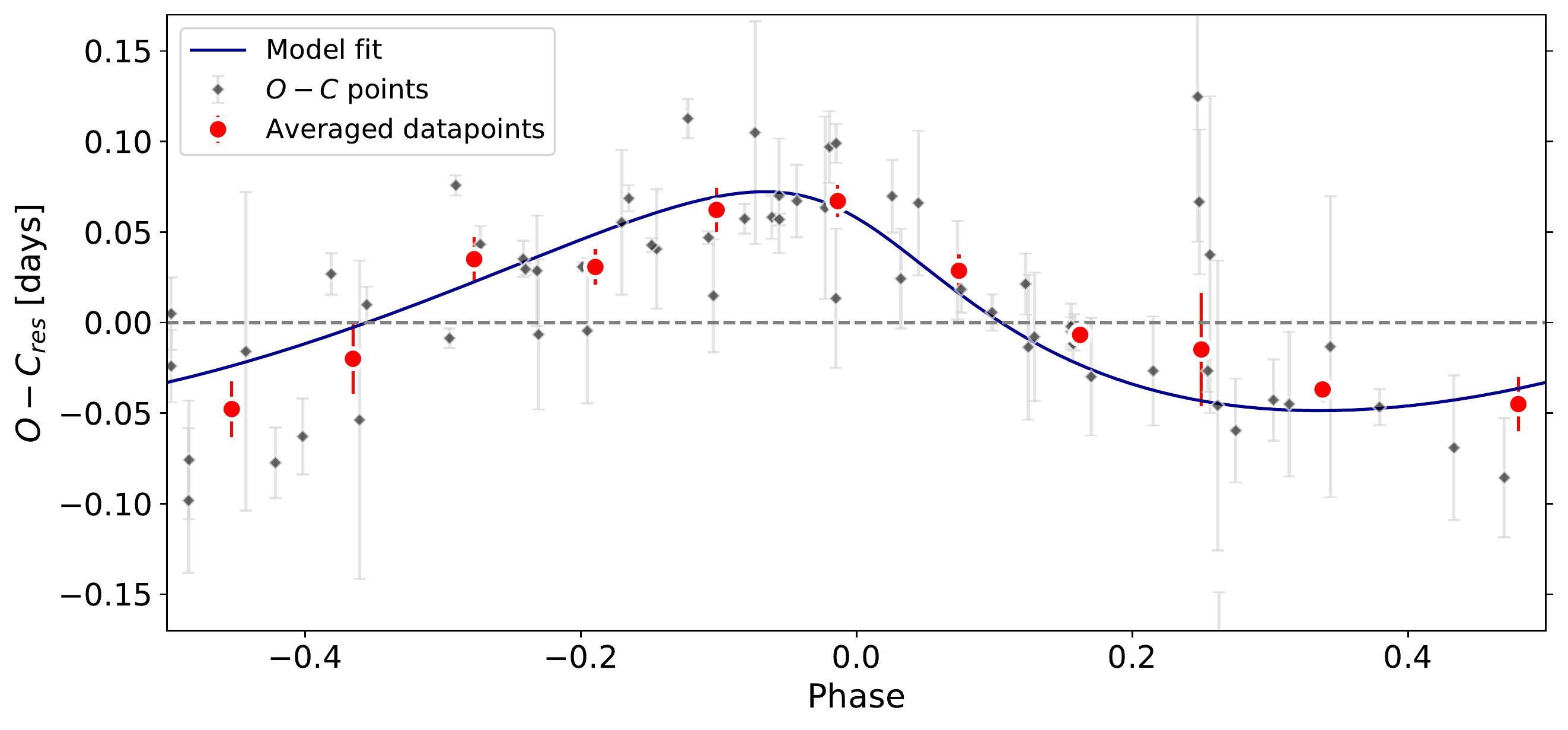}
\caption{Best fit to the residual $O-C$ data of AW~Per calculated by the \texttt{RadVel} software. The grey diamonds represent phase folded $O-C$ residual data, the blue curve the calculated model fit, while the red points show the binned average values of the data points. For the calculation of phases the orbital period of 12078 d has been used.}
\label{fig:AWPer2}
\end{figure}

Similarly to RV data, the LiTE also allows for the detailed estimation of orbital parameters. We must note however, that we did not aim to refine the set of parameters obtained by \cite{Evans2000}, we only investigated this possibility for confirmation purposes. The $O-C$ variation due to the orbital motion can be described as
\begin{equation}
\label{eq:orb}
O-C = \frac{a \sin i}{c} (1-e^2)\left[\frac{\sin (\nu + \omega )}{1+e\cos \nu} - \frac{\sin (\nu_0 + \omega ) }{1+e\cos \nu_0} \right],
\end{equation}
where $a$, $e$, $i$, and $\omega$ refer to the semi-major axis, eccentricity, inclination and longitude of periastron of the orbit of the pulsating star, respectively. In the equation, $\nu$ represents the true anomaly of the pulsator at a given time, while $\nu_0$ denotes the true anomaly at the epoch chosen for the $O-C$ calculation. Based on Eq.~\ref{eq:orb} the semi-amplitude can also be defined:
\begin{equation}
\label{eq:orb}
K = \frac{a \sin i}{c} (1-e^2)
\end{equation}
To fit this profile to the obtained set of $O-C$ residuals, we utilized the \texttt{RadVel} RV fitting toolkit developed by \cite{Fulton2018}, which we appropriately modified using Eq.~\ref{eq:orb} for the fitting of $O-C$ diagrams. Using the Monte Carlo based fitting procedure of the code we obtained a fit plotted in Fig.~\ref{fig:AWPer2} and for which the fitting parameters are listed in Table~\ref{tab:AWPer}. As it is clearly visible from the obtained results, the precision of the $O-C$ diagram based orbital parameters is far worse than the ones obtained by \cite{Evans2000} based on higher precision RV measurements, but our results are consistent with the values obtained previously. It is worth mentioning however, that there is a significant deviation between the orbital period values, as our result for this parameter differs from the one obtained by \cite{Evans2000} ($P = 14594 \pm 324$ d) by more than $7\sigma$. Based on the fact that our photometric dataset covers a longer timeframe than the one covered by RV observations, our result suggests that the orbital period of the binary is indeed significantly shorter than previously thought. To reach a firm conclusion, however, further photometric and spectroscopic observations are required.

\begin{table} 
\begin{tabular}{l r r r}
Parameter & Credible Interval & Max. L. & Units \\
\hline

  $P$ & $12078^{+340}_{-280}$ & $12052$ & d\\
\rule{0pt}{3ex}%
      & $33.07^{+0.93}_{-0.77}$ & $32.99$ & yr\\
\rule{0pt}{3ex}%
  $T_{\rm{conj}}$ & $2422894^{+770}_{-760}$ & $2422915$ &  \\
\rule{0pt}{3ex}%
  $e$ & $0.45^{+0.21}_{-0.2}$ & $0.49$ &  \\
\rule{0pt}{3ex}%
  $\omega$ & $2.38^{+0.47}_{-5.1}$ & $2.6$ & rad \\
\rule{0pt}{3ex}%
  $K$ & $0.051^{+0.01}_{-0.012}$ & $0.053$ & d \\
\rule{0pt}{3ex}%
  $a \sin i$ & $1.66^{+1.14}_{-0.58}\cdot 10^9$ & $1.81\cdot 10^9$ & km \\
\rule{0pt}{3ex}%
 & $11.08^{+7.62}_{-3.88}$ & 12.10 & AU\\
\hline
\end{tabular}
\caption{Orbital elements obtained by fitting the residual $O-C$ diagram of AW~Per using the Monte Carlo based code. The Max. L. columns shows the maximum likelihood estimates of the given parameters.} 
\label{tab:AWPer}
\end{table}

\subsection{Cepheids showing phase jumps/slips}
Of the $O-C$ diagrams we found to be peculiar, 5 could be explained through phase jumps or slips. In case of a phase jump, the phase of the pulsation changes rapidly, while the period remains relatively the same, leading to a stepwise $O-C$ graph. In the case of a phase slip however, the period of the pulsation seems to change rapidly, just to return to the previous value after a given amount of time (period jump-rejump), resulting in a sawtooth-shaped $O-C$ diagram. Such behaviours, which cannot be explained through stellar evolution, have been empirically associated with the Cepheids being binaries, as only some of those were found to exhibit it. However, the complete explanation for this phenomenon is still elusive.

\subsubsection{XZ~Carinae}
XZ~Car is a long period Cepheid which was found to be possibly linked to various open clusters in the past: \cite{Glushkova2015} claimed a possible relation between the Cepheid and the open cluster Ruprecht~93, while \cite{Chen2015} found the membership in ASCC~64 to be more probable, based on  proper motion, age, and instability-strip selection criteria, although none of these associations were marked likely by \cite{Anderson2013} earlier. \cite{Anderson2016} found the Cepheid to exhibit time dependent $\gamma$-velocity changes, indicating the binary nature of the star.

\begin{figure} 
\centering
\includegraphics[width = \linewidth]{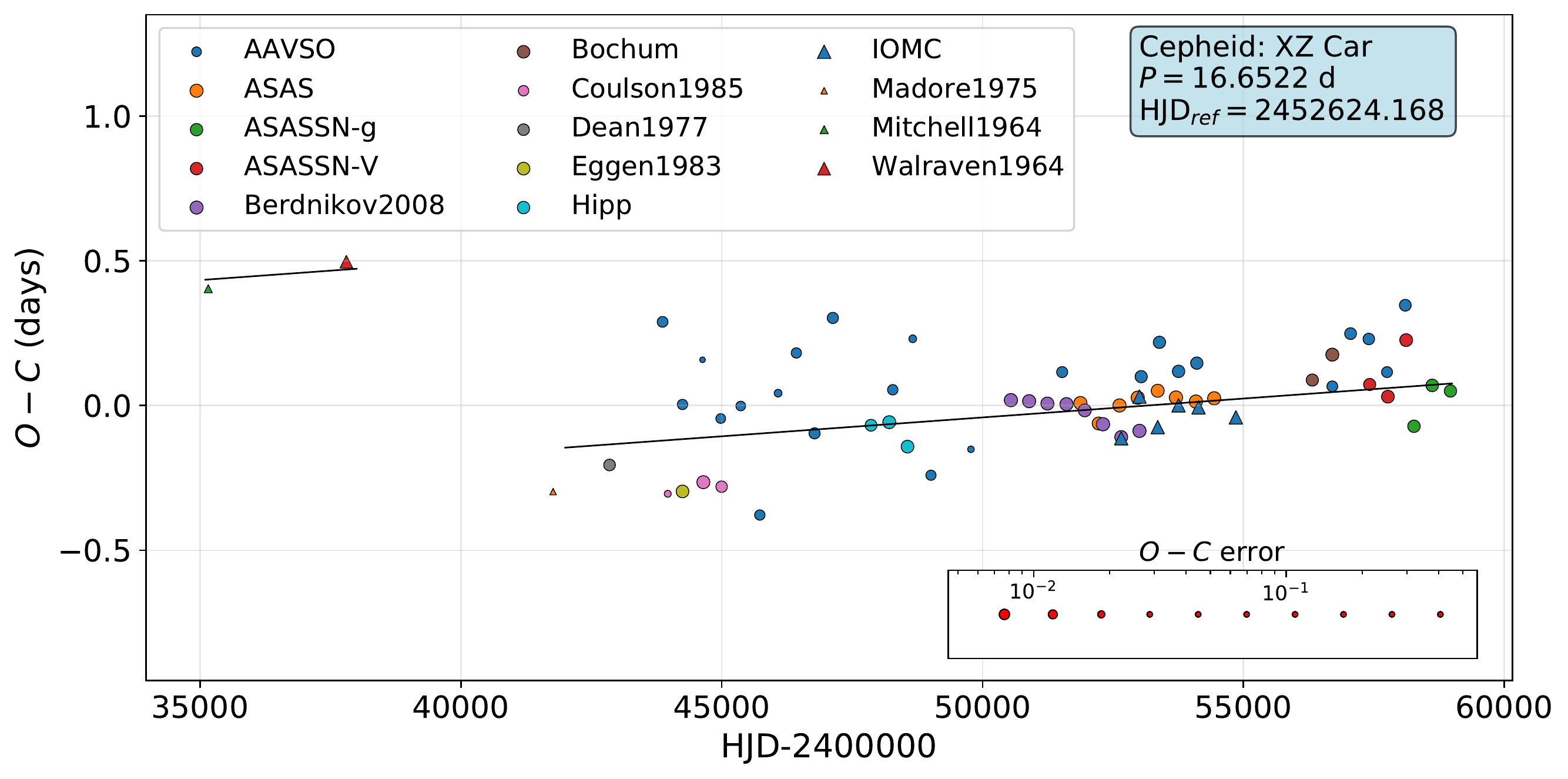}
\caption{$O-C$ diagram of XZ~Car exhibiting a phase jump between HJD~2438000 and 2442000, which altered the phase of the pulsation by 0.77~d.}
\label{fig:XZCar}
\end{figure}

The period changes of this object were not investigated before, despite being a relatively bright star. According to our analysis, the $O-C$ diagram of XZ~Car can only be explained through a phase jump, as the data points available from before HJD 2440000 show a systematic offset from the rest of the data, and no evolutionary signal was visible (Fig.~\ref{fig:XZCar}). The $O-C$ diagram was calculated using the following elements:
$$ C_{\textrm{med}} = 2452624.168 + 16.652232 \cdot E $$

The phase jump occurred between HJD~2438000 and 2442000, and altered the phase of the pulsation by 0.67~d (i.e. 0.040 phase), assuming that the period was unchanged before and after the jump (Fig.~\ref{fig:XZCar}). Apart from the phase jump, no other signal could be extracted that would indicate period change in the Cepheid. We note that due to the low amount of available data before the assumed occurrence of the phase jump this interpretation remains uncertain, however, if true, it could yield an additional empirical link between the presence of a companion star and the phase jump nature of the $O-C$ diagram.

\subsubsection{UX~Carinae}
UX~Car is a short period Cepheid, possibly linked to the open cluster IC~2581 based on its photometric and kinematic properties \citep{Glushkova2015}. However, the latest \textit{Gaia} measurements suggest that the Cepheid is outside the cluster, with the \textit{Gaia} EDR3 distance of the star being 1.52 kpc \citep{Gaia1,GaiaEDR3distances}, while the distance of IC~2581 being 2.45 kpc \citep{Kaltcheva2012}. This association was also marked unlikely by \cite{Anderson2013} previously.

\begin{figure} 
\centering
\includegraphics[width = \linewidth]{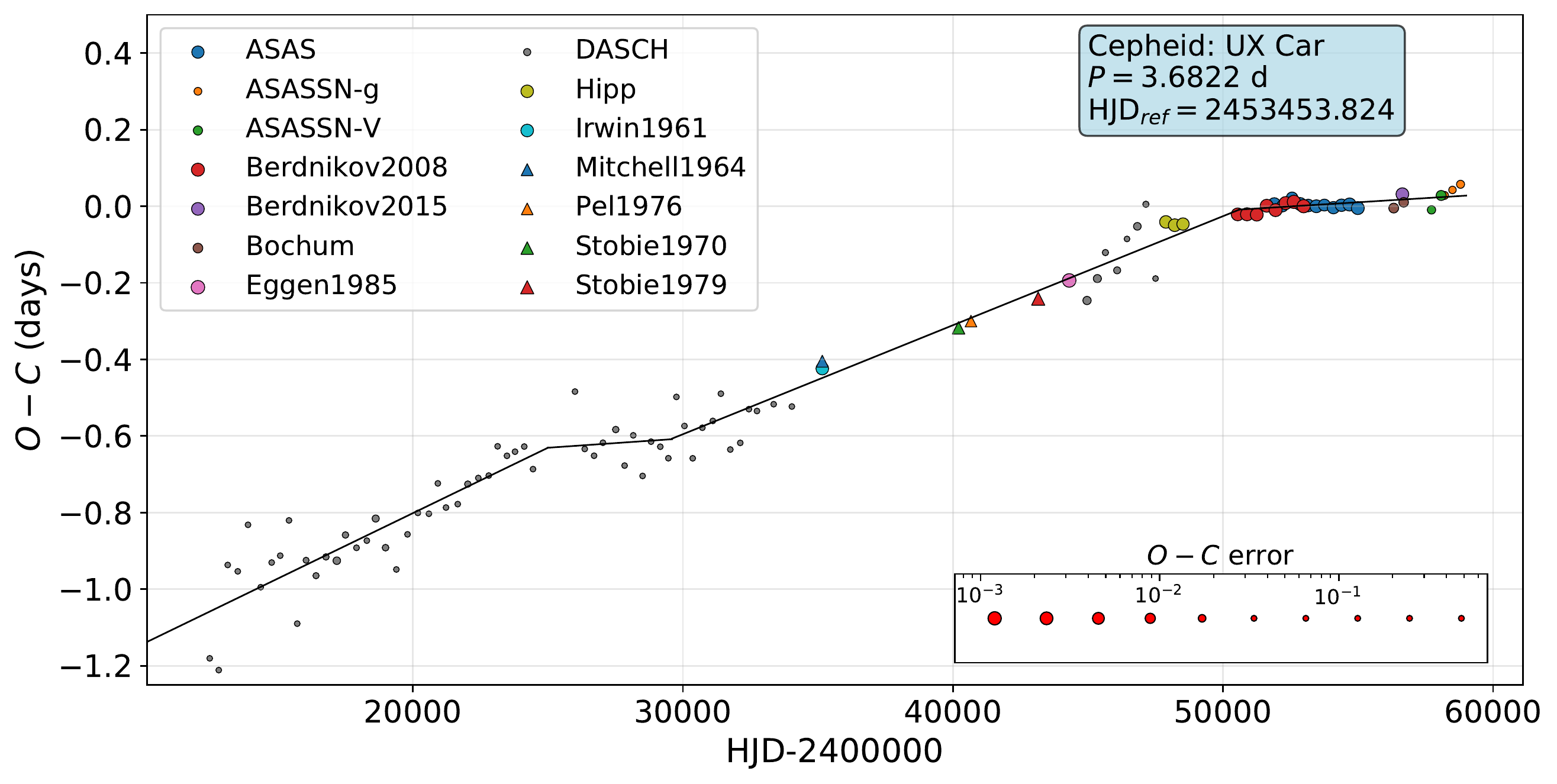}
\caption{$O-C$ diagram of UX~Car, interpreted as a series of period jumps. Three distinct jumps were covered by the observational data, along with the possibility of a fourth one: the first occurred at approximately HJD~2424900, while the second one at HJD~2428900, and the third at HJD~2451000. A fourth period jump could occur at HJD~2457700, but to confirm this, additional photometric observations would be necessary.}
\label{fig:UXCar}
\end{figure}

This Cepheid was not investigated in terms of period changes before. Based on our studies, this variable exhibits rapid period jumps, as it can be seen in Fig.~\ref{fig:UXCar}. The $O-C$ diagram was calculated using the following elements:
$$ C_{\textrm{med}} = 2453453.824 + 3.682171 \cdot E $$

As seen in the figure, the $O-C$ diagram shows a series of period jumps and rejumps, which is best explained by a series of constant period segments:\\
\noindent before HJD 2425000: \hfill $P = 3.6822968 \pm 0.0000129$ d\\
HJD 2425000$-$2429000: \hfill $P = 3.6821750 \pm 0.0000285$ d\\
HJD 2429000$-$2451000: \hfill $P = 3.6822724 \pm 0.0000030$ d\\
after HJD 2451000: \hfill $P = 3.6821861 \pm 0.0000052$ d\\

Since UX~Car was found to be a possible binary by \cite{Kervella2019} based on proper motion studies (although with low probability, due to the large separation), binarity could explain the shape of the $O-C$ diagram, which cannot be interpreted by evolution alone.

\subsubsection{R~Crucis}
R~Cru is a short period Cepheid that has been frequently studied in the past owing to its vicinity to and possible membership in the open clusters NGC~4349 \citep{Majaess2012} or Loden~624 \citep{Anderson2013}. The membership in the former cluster was rejected by \cite{Chen2015}, based on the distance difference between the cluster and the Cepheid and the fact the R~Cru is brighter than the instability strip of the open cluster.

The binary nature of R~Cru was first suspected by \cite{Lloyd1982} based on its $\gamma$-velocity drift, then the presence of the companion was confirmed by HST imaging \citep{Evans2016a} and a possible X-ray emission was also identified at the position of the resolved companion \citep{Evans2016b}. This binary companion was also confirmed by \cite{Kervella2019,Kervella2019a} through finding a slight offset in separation between the \emph{HST} and \emph{Gaia} observation epochs. They also determined the spectral type of R~Cru B, which was found to be compatible with the X-ray radiation.

\begin{figure} 
\centering
\includegraphics[width = \linewidth]{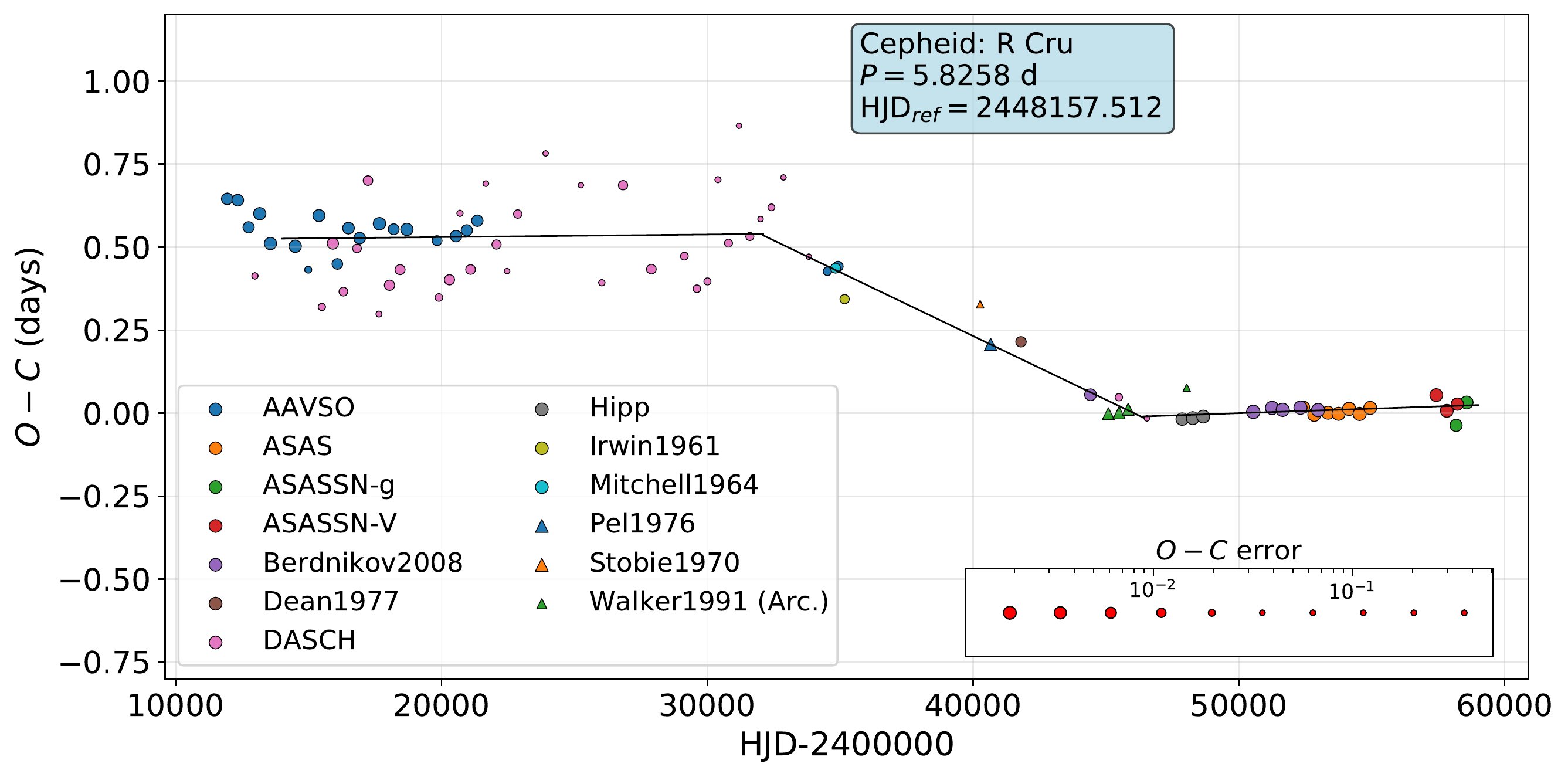}
\caption{$O-C$ diagram of R~Cru exhibiting a period jump at HJD 2433000, which modified the period of the Cepheid until HJD~2446000. The jump--rejump nature of the $O-C$ diagram provides another indication for the existence of the binary component.}
\label{fig:RCru}
\end{figure}

Despite being relatively bright, no $O-C$ diagram has been constructed yet for this Cepheid. According to our analysis, R~Crucis underwent a period jump approximately at HJD~2433000 and pulsated with a modified period until approximately HJD~2446000 (Fig.~\ref{fig:RCru}):\\
\noindent before HJD 2432900: \hfill $P = 5.8257576 \pm 0.0000141$ d\\
HJD 2432900$-$2446100: \hfill $P = 5.8255291 \pm 0.0000109$ d\\
after HJD 2446100: \hfill $P =  5.8257880 \pm 0.0000061$ d\\

The presented $O-C$ diagram was calculated using the following elements: 
$$ C_{\textrm{med}} = 2448157.512 + 5.825771 \cdot E $$

After the phase jump the period of the Cepheid returned to a value slightly longer than before, which suggests that the period of the Cepheid is increasing, however this could only be determined by observing another period jump. Nevertheless, the jump--rejump structure of the $O-C$ diagram could be linked to the presence of a companion, as evolutionary changes alone cannot explain this behaviour.

\subsubsection{DT~Cygni}
DT~Cyg is a short period Cepheid which was frequently observed in the past decades. The $O-C$ diagram of the Cepheid constructed by \cite{Szabados1991} showed peculiar period jump--rejump behaviour. This Cepheid was also suggested to be in a binary system, first by \cite{Leonard1986}, then by \cite{Szabados1991} based on RV measurements, however, the orbital motion is yet to be confirmed.

\begin{figure} 
\centering
\includegraphics[width = \linewidth]{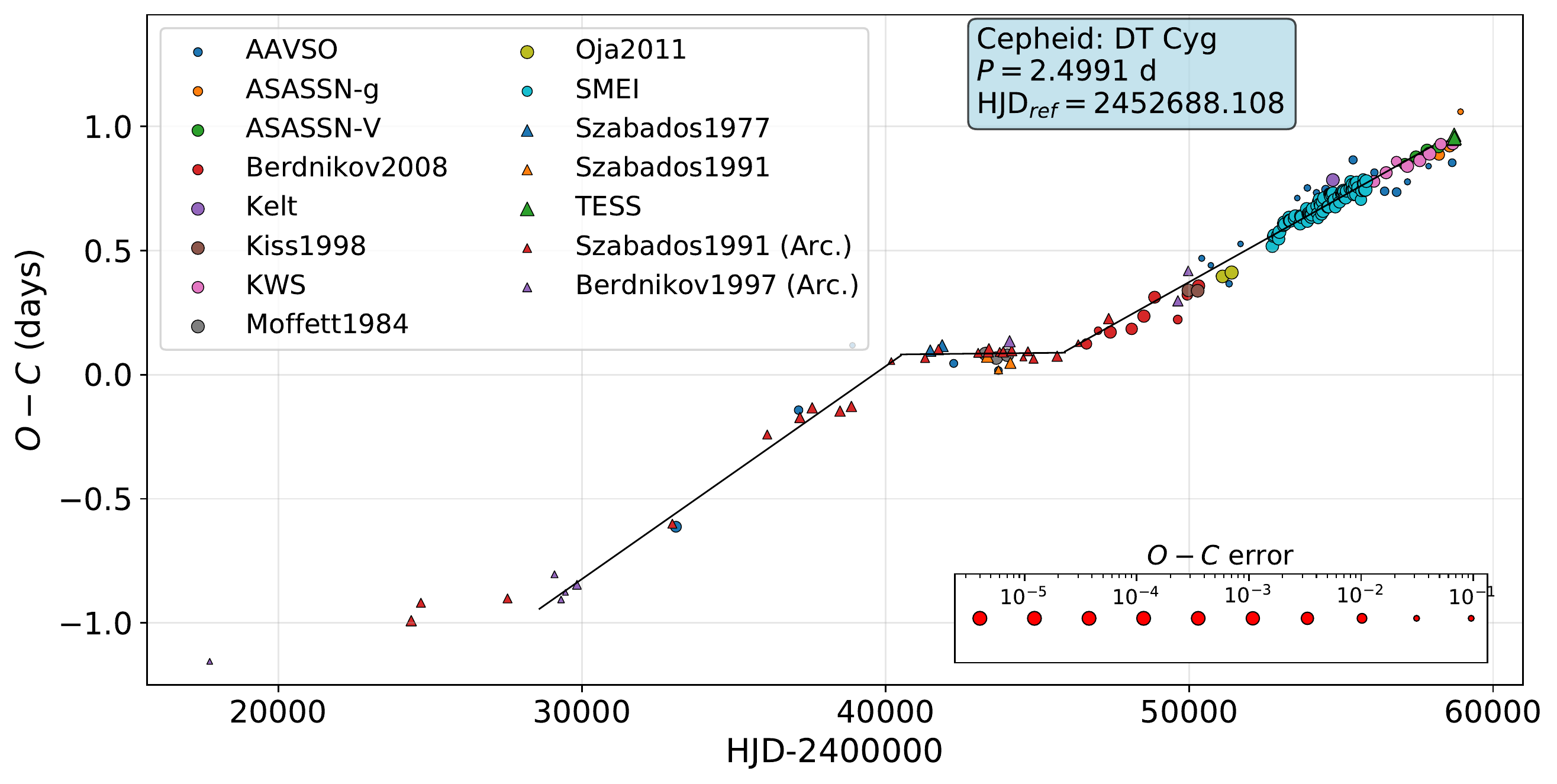}
\caption{$O-C$ diagram of DT~Cyg, exhibiting a clear case of period jump.}
\label{fig:DTCyg}
\end{figure}

By analysing the obtained $O-C$ diagram of DT~Cyg we confirm the previous statement by \cite{Szabados1991}, that a second rapid period change occurred near HJD~2446000 (Fig.~\ref{fig:DTCyg}). Our $O-C$ diagram was calculated using the following elements:
$$ C_{\textrm{med}} = 2452688.108 + 2.499063 \cdot E $$

 The pulsation of DT~Cyg can be characterized with the following periods:\\
\noindent before HJD~2441300: \hfill $P = 2.4992250 \pm 0.0000093$ d\\
HJD~2441300$-$2445900: \hfill $P = 2.4990380 \pm 0.0000116$ d\\
after HJD~2445900: \hfill $P = 2.4992363 \pm 0.0000025$ d\\

It is not possible to determine how the period changed before HJD~2428600 based on the current diagram. The peculiar structure of the $O-C$ diagram could be linked to the presence of a companion, since evolutionary changes alone cannot explain it. Moreover, as it can be seen from the values above, the average pulsation period of DT~Cyg became shorter after the rejump, which suggests that the star belongs to the second crossing Cepheids.

\subsubsection{BN~Puppis}

BN~Pup in an intermediate period Cepheid which was not investigated in terms of period changes before. The Cepheid was suspected to be connected to NGC~2533, but this turned out to be unlikely based on the large distance and age differences \citep{Havlen1976, Anderson2013}. The binary nature of the object was not investigated before, with only limited amount of RV data being available \citep{Coulson1985, Pont1994, Storm2011}.

\begin{figure} 
\centering
\includegraphics[width = \linewidth]{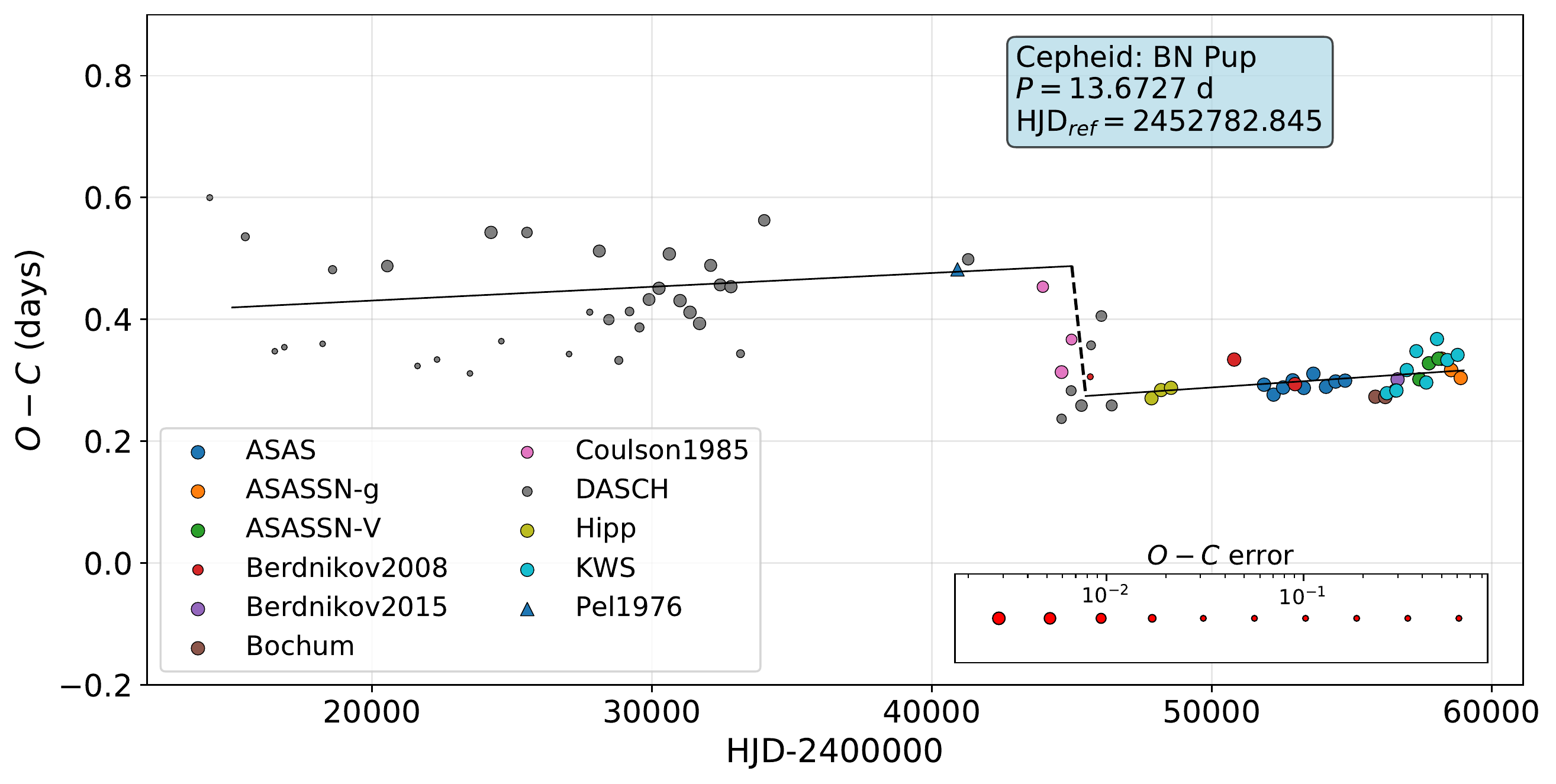}
\caption{$O-C$ diagram of BN~Pup, with a phase jump of ~0.2 d visible at approximately HJD~2444500 which indicates binarity of the Cepheid.}
\label{fig:BNPup}
\end{figure}

Based on the compiled $O-C$ diagram plotted in Fig.~\ref{fig:BNPup}, we have some indication that the Cepheid might have a companion. The phase jump in the case of this Cepheid is less ambiguous than in other cases since it is detected in data from a single source (DASCH photometry) and it is also supported by the precise photometric measurements of \cite{Pel1976}. The presented diagram was calculated using the following elements:
$$ C_{\textrm{max}} = 2452782.845 + 13.672693 \cdot E  $$

The phase jump occurred at approximately HJD~2444500 and altered the phase of the pulsation with $\sim$0.2~d. The values of the pulsation period before and after the phase jump are in agreement within errors. To determine whether a long period signal can be extracted from the RV data, we reanalysed the measurements of \cite{Coulson1985}, \cite{Pont1994}, and \cite{Storm2011}. According to the Fourier decomposition of the signal, a period of $\sim$4300~d can be extracted, but the sparseness and the uncertainty of the RV data does not allow us to determine whether it is truly connected to the binary nature of the Cepheid. The phase jump exhibited by the $O-C$ diagram could suggest that BN~Pup is possibly a member in a binary system, if one assumes an empirical correspondence between the two, although further RV measurements are necessary to validate this assumption.

\subsection{Overtone Cepheids}
The remaining two Cepheids that were found to exhibit peculiar $O-C$ diagrams show erratic or semi-erratic period changes, which, to our current best understanding, can only be explained through a set of linear segments (consecutive intervals of random, but otherwise constant-period pulsations) and is a characteristic of overtone pulsators. During the observed timeframe, none of these Cepheids showed recurrence in the change of their pulsation period.

\subsubsection{X~Lacertae}
X~Lac is a short period Cepheid which is possibly connected to the open cluster FSR~384 \citep{Glushkova2015} (although this association was marked as unlikely by \cite{Anderson2013} before). The period changes of this Cepheid were extensively studied by \cite{Evans2015}, finding that it shows an erratic period change. Apart from the erratic period change, the low pulsation amplitude also points towards the overtone nature of the pulsation \citep{Storm2011}, which makes this object interesting, as only a few overtone pulsators are known with such a long period.

\begin{figure} 
\centering
\includegraphics[width = \linewidth]{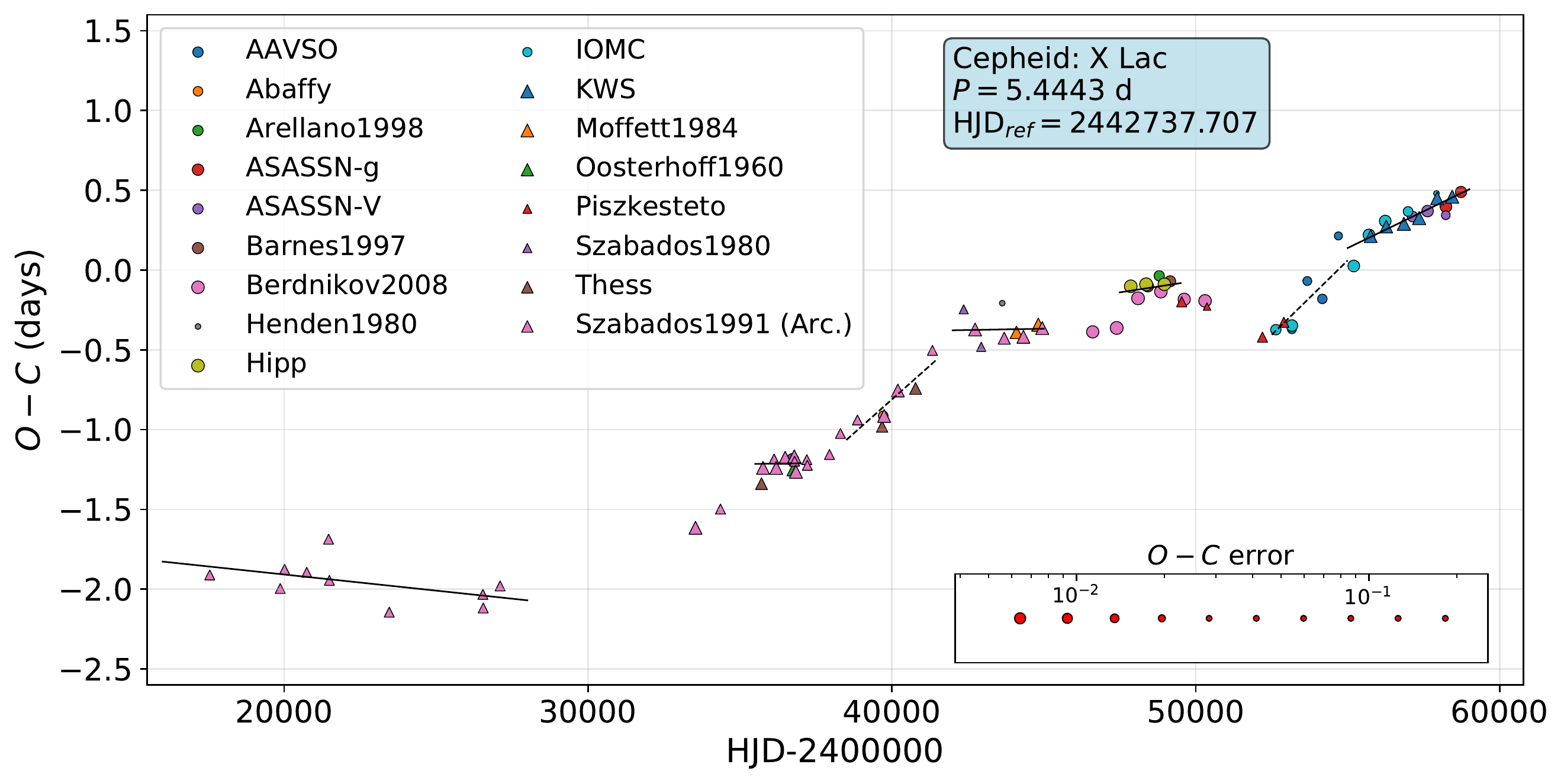}
\caption{$O-C$ diagram of X~Lac exhibiting an erratic behaviour which can best be described with a series of rapid period changes.}
\label{fig:XLac}
\end{figure}

Our analysis of the available photometric data yielded very similar results to those presented by \cite{Evans2015}: the resulting $O-C$ diagram is best fitted with a series of linear segments as shown in Fig.~\ref{fig:XLac}. The presented $O-C$ diagram was calculated with the following elements:
$$ C_{\textrm{med}} = 2442737.707 + 5.444322 \cdot E $$

According to our analysis, the $O-C$ diagram of X~Lac can be best described with the following periods:\\
\noindent before HJD 2428000: \hfill $P = 5.4442119 \pm 0.0000659$ d\\
HJD 2435500$-$2437000: \hfill $P = 5.4445175 \pm 0.0001767$ d\\
HJD 2438500$-$2441500: \hfill $P = 5.4456094 \pm 0.0002935$ d\\
HJD 2442000$-$2445000: \hfill $P = 5.4443820 \pm 0.0000715$ d\\
HJD 2447500$-$2449500: \hfill $P = 5.4444422 \pm 0.0001904$ d\\
HJD 2452500$-$2455000: \hfill $P = 5.4451025 \pm 0.0003423$ d\\
after HJD 2455000: \hfill $P = 5.4448548 \pm 0.0000447$ d\\

Our analysis shows that the period change sequence of X~Lac is indeed erratic throughout the observed timeframe, which supports the assumption that this Cepheid is an overtone pulsator.

\subsubsection{EU~Tauri}

EU~Tau is a short period Cepheid and an overtone pulsator \citep{Gieren1990}. The binary nature of the object was first suspected based on RV data \citep{Gorynya1996}, however a later study by \cite{Evans2015} found this unlikely. The period changes of EU~Tau were investigated by \cite{Berdnikov1997} and \cite{Evans2015}, finding first a linear, then erratic period change, respectively.

\begin{figure} 
\centering
\includegraphics[width = \linewidth]{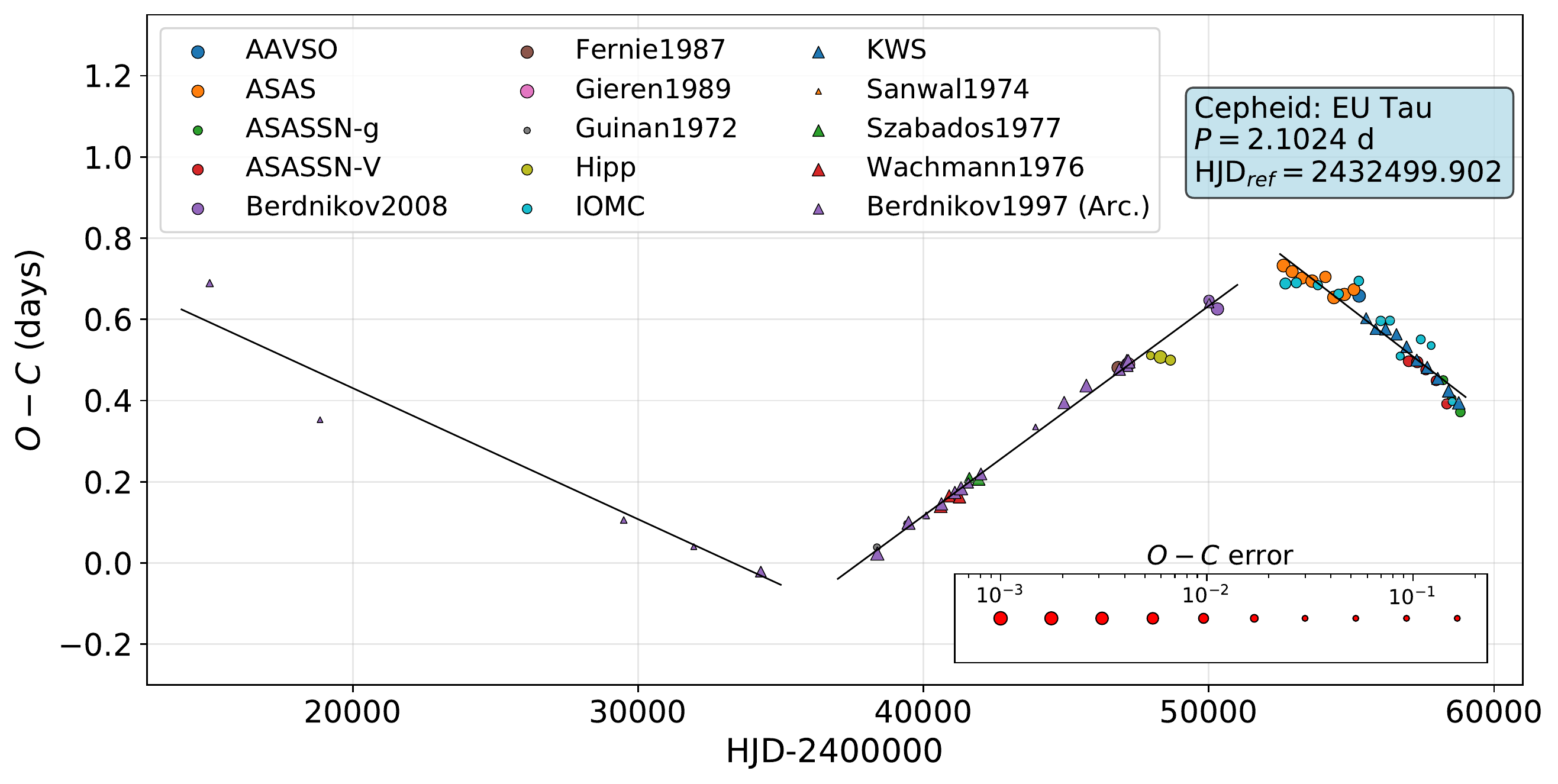}
\caption{$O-C$ diagram of EU~Tau, showing two rapid period changes, between which the pulsation of the star can be described with constant period values.}
\label{fig:EUTau}
\end{figure}

The obtained $O-C$ diagram (Fig.~\ref{fig:EUTau}) was calculated using the following elements:
$$ C_{\textrm{med}} = 2432499.902 + 2.10241 \cdot E. $$

According to our analysis, which matches with the finding of \cite{Evans2015}, the $O-C$ diagram can be best described as three constant period segments:

\noindent before HJD 2435000: \hfill $P = 2.1023336 \pm 0.0000028$ d\\
HJD 2437000$-$2451000: \hfill $P = 2.1025187 \pm 0.0000017$ d\\
after HJD 2452500: \hfill $P = 2.1022962 \pm 0.0000047$ d\\

Since the transitions between these segments were not covered by observations, their nature cannot be established. To determine whether the pulsation periods are truly erratic, or if there is any connection between them, further observations are required.

\section{Collective analysis on period change rates and fluctuations}
\label{sec:4}

\subsection{Colour-magnitude diagram and period change rates}
Apart from the Cepheids detailed in the previous sections, we have obtained $O-C$ diagrams for 132 additional Cepheids (see Table~\ref{tab:allceps} for the complete list), which all show a parabolic shape, thus dominated by the evolution of the variable star. As a result, our Cepheid sample contains 141 stars that exhibit clear evolutionary changes in their $O-C$ diagrams, including those that were termed peculiar based on additional features in their $O-C$ diagrams in Sect~3.1.

The recent advancement in astrometry owing to the measurements of the {\em Gaia} space probe allows for the construction of the colour-magnitude diagram (CMD) for the current set of Cepheids with an unprecedented accuracy. Figure~\ref{fig:CMDcomp} shows the CMD based on the {\em Gaia}~DR2 and {\em Gaia}~EDR3 data \citep{Gaia1,Gaia2,GaiaEDR3,Gaia3,Gaia4} as a comparison. Since neither of the data releases contained measurements of the Galactic extinction for all the Cepheids, we applied the attenuation correction on the individual photometric measurements based on the colour-excess values listed in the DDO database of Galactic Classical Cepheids\footnote{\url{http://www.astro.utoronto.ca/DDO/research/cepheids/cepheids.html}}. For the Cepheids which were not listed on the website, or scattered far from the rest of the variables on the CMD due to the false value of the colour excess (KL~Aql, V898~Cen) we used the more recent measurements  \citep{Schmidt2015, Groenewegen2020}, but in the case of V898~Cen the resulting position was still slightly different from the rest of the dataset. For the additional Cepheids that have no published colour-excess values at all (EV~Cir, V5738~Sgr, GP~Mus, V2744~Oph, CE~Pup, DU~Pyx and V520~Vel) we assumed the extinction values inferred by \cite{Anders2019}. The colour excess values were then converted to the {\em Gaia} passband system using the scaling given  in \cite{Casagrande2018}. 

As seen in Fig.~\ref{fig:CMDcomp}, the data from EDR3 show a much smaller scatter and exhibit a much clearer trend for the pulsation period of the Cepheids, which demonstrates well the improvement of the {\em Gaia} data compared to the previous data releases. On the right panel of Fig.~\ref{fig:CMDcomp} we attempted to draw the limiting lines of the strip populated by the Cepheids; for the approximate alignment of these lines we omitted all Cepheids with uncertain or non-published extinction mentioned above. The resulting lines are notably not parallel, which indicates that for longer period Cepheids the possible range of colour index values is broader, which means that the instability \emph{strip} of Cepheids could in fact be an instability \emph{wedge}, as proposed earlier by \cite{Fernie1990a}. This also matches the results previously obtained from the observations of IUE for a smaller sample of Cepheids \citep{Evans1992b}. To examine this possibility even further, more accurate extinction values would be required, which are planned to be published with the complete third data release of {\em Gaia} in 2022.

\begin{figure} 
\centering
\includegraphics[width = \linewidth]{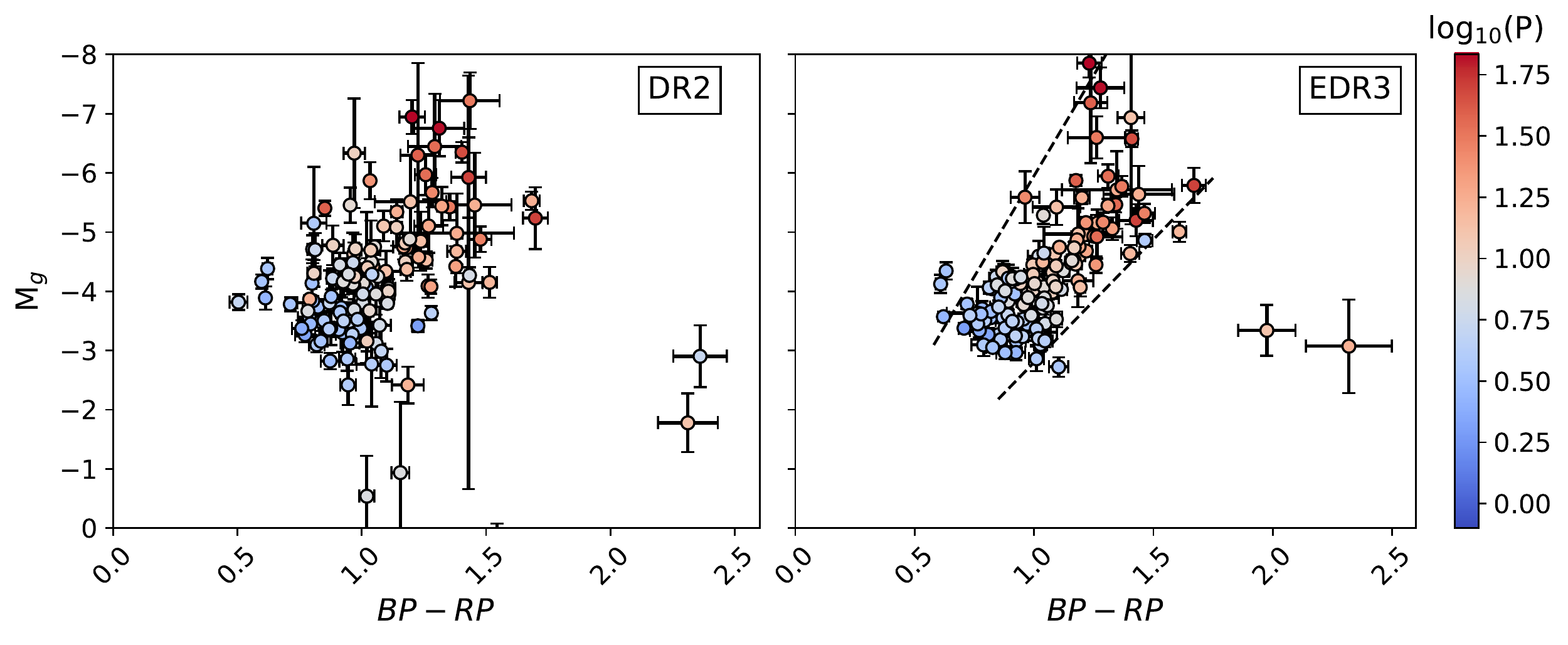}
\caption{Colour-magnitude diagrams of the investigated Cepheids calculated from the {\em Gaia} DR2 (left) and {\em Gaia} EDR3 (right) datasets. The dashed lines show the assumed edges of the instability strip. The error bars show the uncertainties arising due to the parallax and the reddening errors.}
\label{fig:CMDcomp}
\end{figure}

\begin{figure} 
\centering
\includegraphics[width = \linewidth]{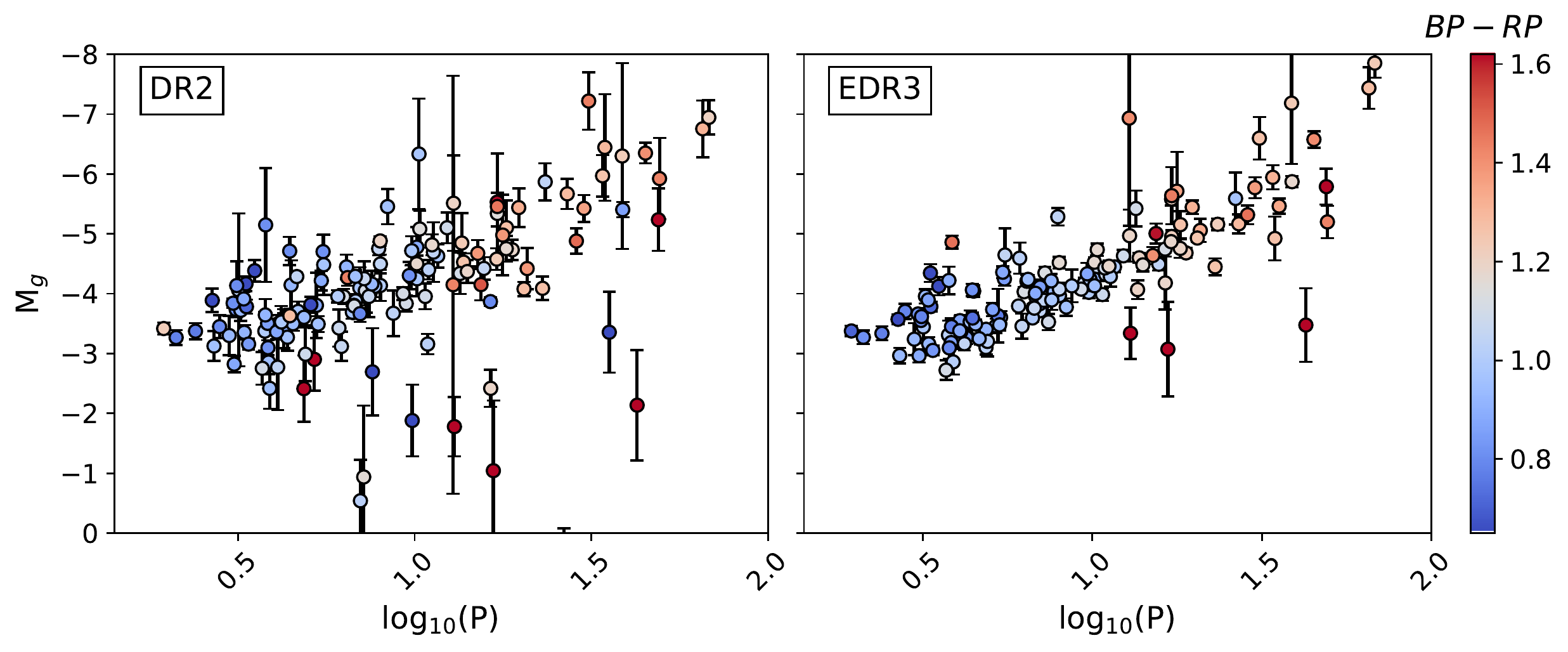}
\caption{Period-luminosity relation of the investigated Cepheids, calculated from the {\em Gaia} DR2 (left) and {\em Gaia} EDR3 (right) datasets. The individual points are coloured by the dereddened {\em Gaia} $BP-RP$ values. The right panel demonstrates well the improvement of quality in the Cepheid distances between the two data releases.}
\label{fig:CMDcomp}
\end{figure}

By fitting the $O-C$ diagrams of the investigated Cepheids with second order polynomials, along with a sinusoidal modulation superimposed when statistically motivated, and then applying the method described by \cite{Sterken2005} we could determine the rates of period change for each of the variable stars. Our results and conclusions on period change rates essentially match those presented by \cite{Turner2006}: the values inferred by us are well explained by the evolutionary models discussed in that work and the ratio of Cepheids showing positive and negative period changes is also identical, if one considers the possible selection effects (70\% of our sample Cepheids show positive period changes, while this value was 67\% in \cite{Turner2006}). The deviations from evolutionary models we find in our sample are also very similar to the previous results: several short period Cepheids exhibit a period change rate that fall between the value ranges expected for first and third crossing Cepheids, while several intermediate period Cepheids show a period change rate that is smaller than predicted by evolutionary models.

As it was discussed by \cite{Neilson2012}, the inclusion of convective core overshooting and enhanced mass loss can explain these deviations. Moreover, they showed that the difference between the ratio of stars showing positive and negative period changes for observations and predictions (where the unmodified evolutionary models estimate significantly larger fraction of positive period change Cepheids, namely 85\%) can also be explained through mass loss, although they note that there is currently no explanation for large mass loss rate they found to be required. Complementary to this work, \cite{Miller2020} investigated how rotation and convective core overshooting alter the distribution of period change rates, finding that they do not assert significant influence on then, despite being important parameters for the evolution of Cepheids.

\begin{figure} 
\centering
\includegraphics[width = \linewidth]{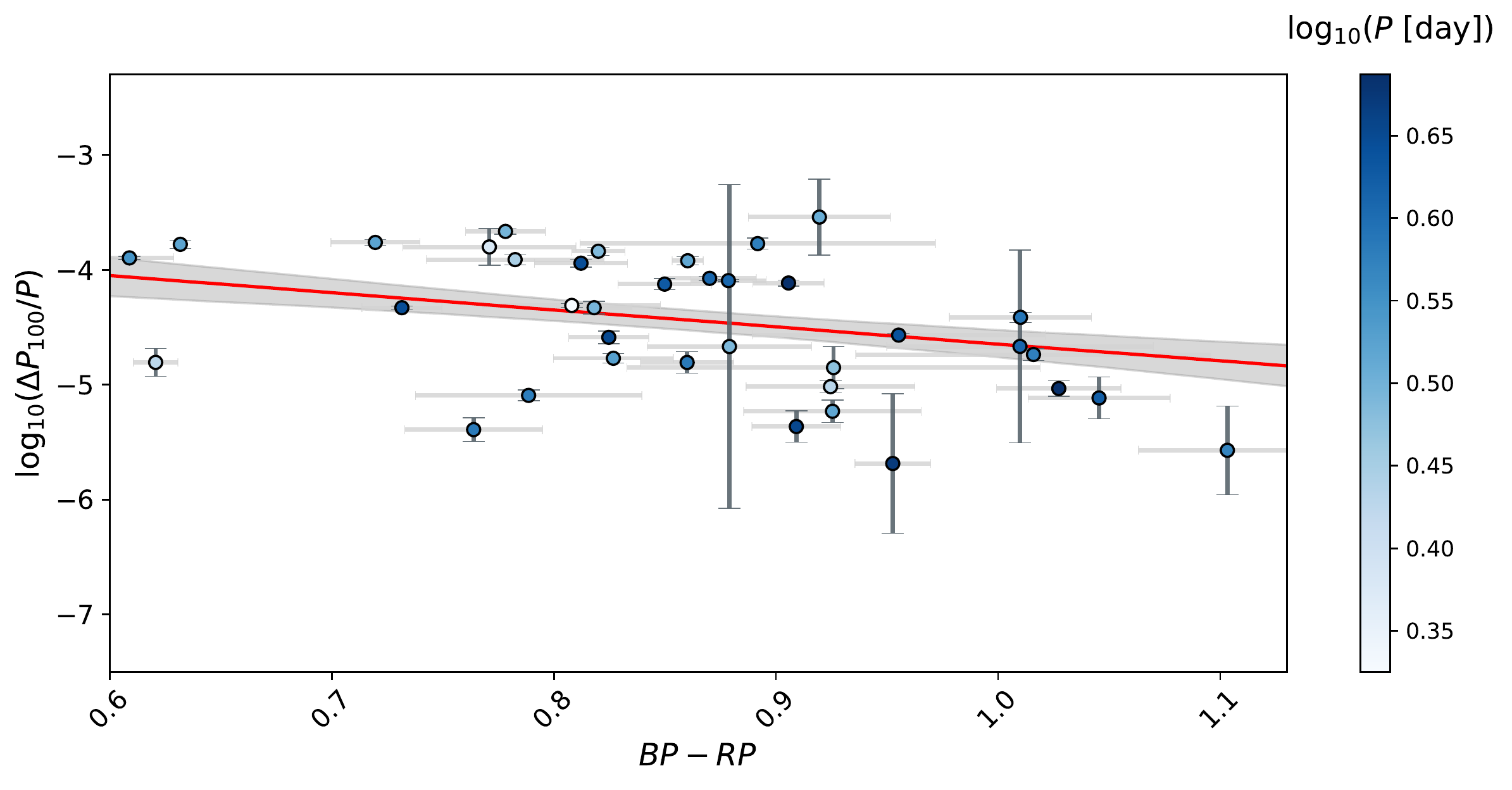}
\caption{Correlation plot of the colours and period changes of short period Cepheids (pulsation period range of 2-5 days). The red line shows the best linear fit to the whole dataset, while the grey shaded area shows the $1\sigma$ quantile of the fit. Period changes decrease within errors towards the red edge of the instability strip.}
\label{fig:Pdotcolourfit}
\end{figure}

As it was noted by \cite{Turner2006}, the period change rate could be an important probe for the estimation of stellar properties. To test this observationally, we correlated the period change rates of Cepheids to their colours for a moderate range of pulsation periods (between 2 and 5 days) as seen in Fig.~\ref{fig:Pdotcolourfit}. The period range was chosen based on the CMD, as the relevant effect was visible primarily for short period Cepheids, while for longer periods the scatter on the CMD was too large for detecting a meaningful effect. As seen in the plot, bluer Cepheids exhibit slightly larger period change rates, which decrease toward the red edge of the instability strip. The individual points are coloured by the pulsation period length, which show no significant correlation with either the period change rate or the colour within the considered period range. This agrees well and observationally validates the statement in \cite{Turner2004}, in which they note that at a given pulsation period, Cepheids at the high-temperature edge of the instability strip should exhibit larger rate of period change than those at the low-temperature edge, due to the former having larger masses. 

This demonstrates that the period change rate can indeed provide a valuable observational probe for the determination of physical parameters. In addition to the pulsation period, which can be used to estimate both the ``vertical'' and the ``horizontal'' position of the Cepheid within the instability strip, and the corresponding physical parameters owing to the period-luminosity and period-colour relations (assuming an underlying mass-luminosity relation and a pulsation model, see, e.g., \citealt{Beaulieu-2001}), the period change rate can independently probe the ``horizontal'' direction. Since the period and period change rate values are not tightly correlated due to the various crossing modes, mass loss, and convective core overshooting rates, this correlation is not bound to the pulsation period, which is also well shown in Fig.~\ref{fig:Pdotcolourfit}. Henceforth, the period change rate could be used to enhance the precision of colour and temperature estimation of Cepheids when coupled with the period values. Further investigation of this correlation is beyond the scope of this work, however we note that it would be worthwhile to analyse to what extent can it be used for the inference of physical properties. Combined with the new data from {\em Gaia} it could provide a simple way to constrain the effective temperature of the variables compared to the conventional spectroscopic method, yet it could prove to be more precise than the current, solely colour based empirical relations.

Owing to the long temporal coverage of the newly calculated $O-C$ diagrams one can not only calculate the period change rate, but its acceleration as well: in such cases the resulting $O-C$ diagrams systematically deviate from the parabolic fit, since their complete description  would require higher-order polynomials, as it was also noted in previous works \citep{Szabados1983, Fernie1984, Turner2006}. In our sample, most of the long period Cepheids exhibit such $O-C$ diagrams (most notably, EV~Aql, S~Vul and SV~Vul). Naturally, all Cepheids are expected to show such high order terms in their $O-C$ diagrams, but due to the short observed timeframes these terms are usually negligible for Cepheids that exhibit smaller period change rates.

\begin{figure} 
\centering
\includegraphics[width = \linewidth]{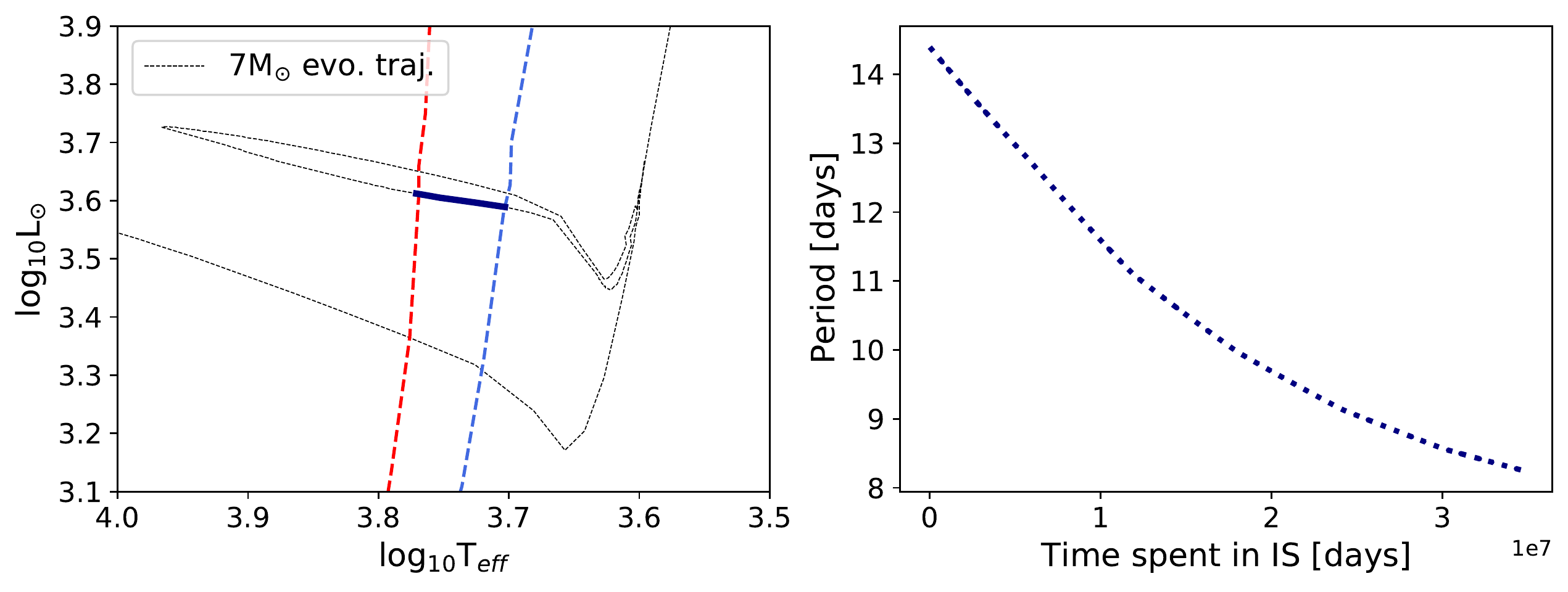}
\caption{Period values estimated for the second crossing of a 7M$_\odot$ star at different times (right panel), showing a curved evolution indicating a non-linearly changing period, with the left panel showing the corresponding evolutionary trajectory (with the important segment highlighted).}
\label{fig:evotraj}
\end{figure}

To visualize this expectation we attempted to recover the higher order period change from theoretical models: by adopting the evolutionary trajectory models and instability strip edges from \cite{Georgy2013} and \cite{Anderson2016} respectively, and the period-mass-luminosity-temperature relations of \cite{DeSomma2020}, we calculated how the pulsation period would change for a single star due to stellar evolution. An example of the results is shown in Fig.~\ref{fig:evotraj}: the theoretical models also yield non-linearly changing period, which validates the expectation for higher order terms in the $O-C$ values. Currently, the interpretation of these higher order terms is not yet complete, but they will also yield additional valuable information about the pulsation of these stars similarly to the period change rates.

\subsection{Period fluctuations}
\label{sec:fluct}
\begin{figure} 
\centering
\includegraphics[width = \linewidth]{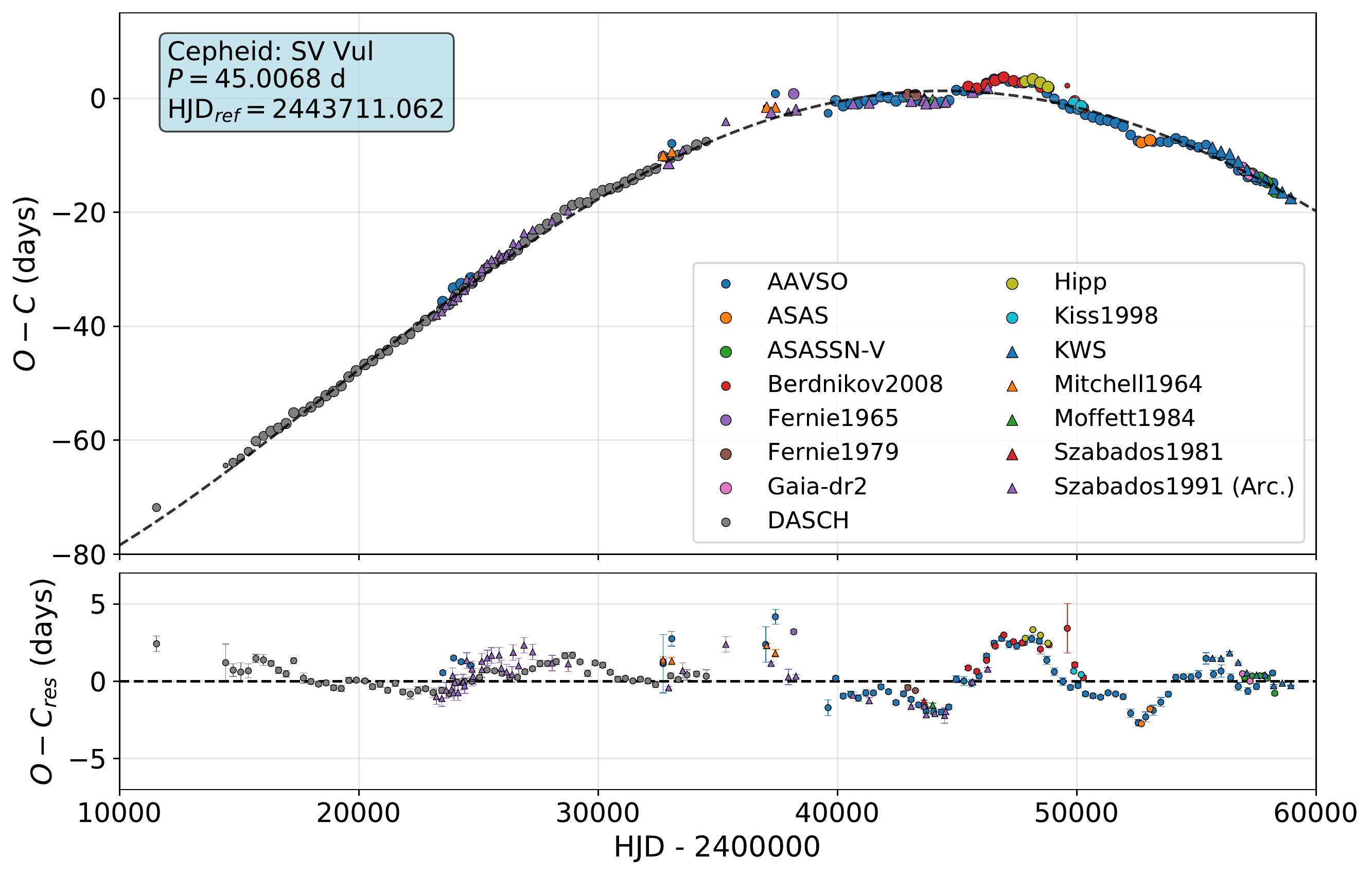}
\caption{$O-C$ diagram of SV~Vul (top panel), showing a long-term change corresponding to stellar evolution, with superimposed period fluctuations, and the residual $O-C$ diagram (bottom panel) after subtracting the evolutionary changes.}
\label{fig:SVVul}
\end{figure}

Apart from examining the overall rate of period change for the Cepheids in our sample, the $O-C$ diagrams also made it possible to investigate random fluctuations present in the pulsation. An example for such fluctuation signal can be seen in the $O-C$ diagram of SV Vul (Fig.~\ref{fig:SVVul}): it shows a long-term trend corresponding to the evolution of the Cepheid and a quasi-periodic wavelike signal superimposed, with varying shape. Such wavelike signals cannot be explained by LiTE, as their amplitudes are too large compared to their period, and the shape of the waves also changes in time. The physical explanation for such modulations is that random fluctuations in the pulsation period accumulate and cause wavelike signals of varying shapes. For long, this was the only classical Cepheid for which fluctuations were detected \citep{Turner2009}. Recently, \cite{Rodriguez-Segovia2021} also found that five additional long period Cepheids in the LMC reliably show such large amplitude fluctuations, highlighting their ubiquity. It was expected that fluctuations are present in most Cepheid variables as well, but detection would require continuous space-based photometry, where consecutive epochs can be observed \cite{Turner2010}. Multiple studies attempted to detect and quantify fluctuations this way, as we discussed in Sect.~\ref{Sec:1}.

As we will show below, such period fluctuations are more common among Cepheids than previously thought. It has been shown that longer period variables, like RV~Tauri or Mira stars, can exhibit long term quasi-periodic waves in their $O-C$ diagrams due to the period fluctuations, which becomes more prominent for longer pulsation periods \citep{Percy1997, Percy1999, Molnar2019}, however, no such conclusions were drawn for classical Cepheids. In the sample of stars we investigated, we found that 51 Cepheid showed wavelike signals in the residual $O-C$ diagrams that most probably originated from fluctuations, apart from the Cepheids that were termed `peculiar' in the previous sections. To select these Cepheids from our sample, we performed F-tests on their $O-C$ diagram fits to analyse the residuals of the models and to check whether they indicate that the use of more complex models (i.e. which also fit for an assumed wavelike modulation) are validated. We compared three different models for the test: the constant period model (linear $O-C$ diagram), the linear period change model (parabolic $O-C$ diagram) and the linear period change plus wavelike modulation (superposition of a parabolic and a sinusoidal change), in which case we simultaneously fit for both the evolutionary changes and the modulation. Table~\ref{tab:ftest} of the Appendix show the results obtained for the different model pairs. We included Cepheids in the fluctuating sample when the F-test conducted for the comparison of the parabolic and parabolic plus wave fits yielded an F-statistics larger than 10, which means that the more complex model was favoured at a significance level of 0.99 (with the difference between the degrees of freedom of the compared models being $DF = 3$). To remove possible contaminations, or Cepheids for which the fitting code yielded erroneous results, we rejected every star where the period of the sinusoidal fit was either shorter than 3000 days (8-10 times the folding time used for the construction of the $O-C$ diagrams, to avoid sampling effects) or longer than five quarters of the observed time frame (in which case the sinusoidal waveform overfitted the parabolic change).

\begin{figure} 
\centering
\includegraphics[width = \linewidth]{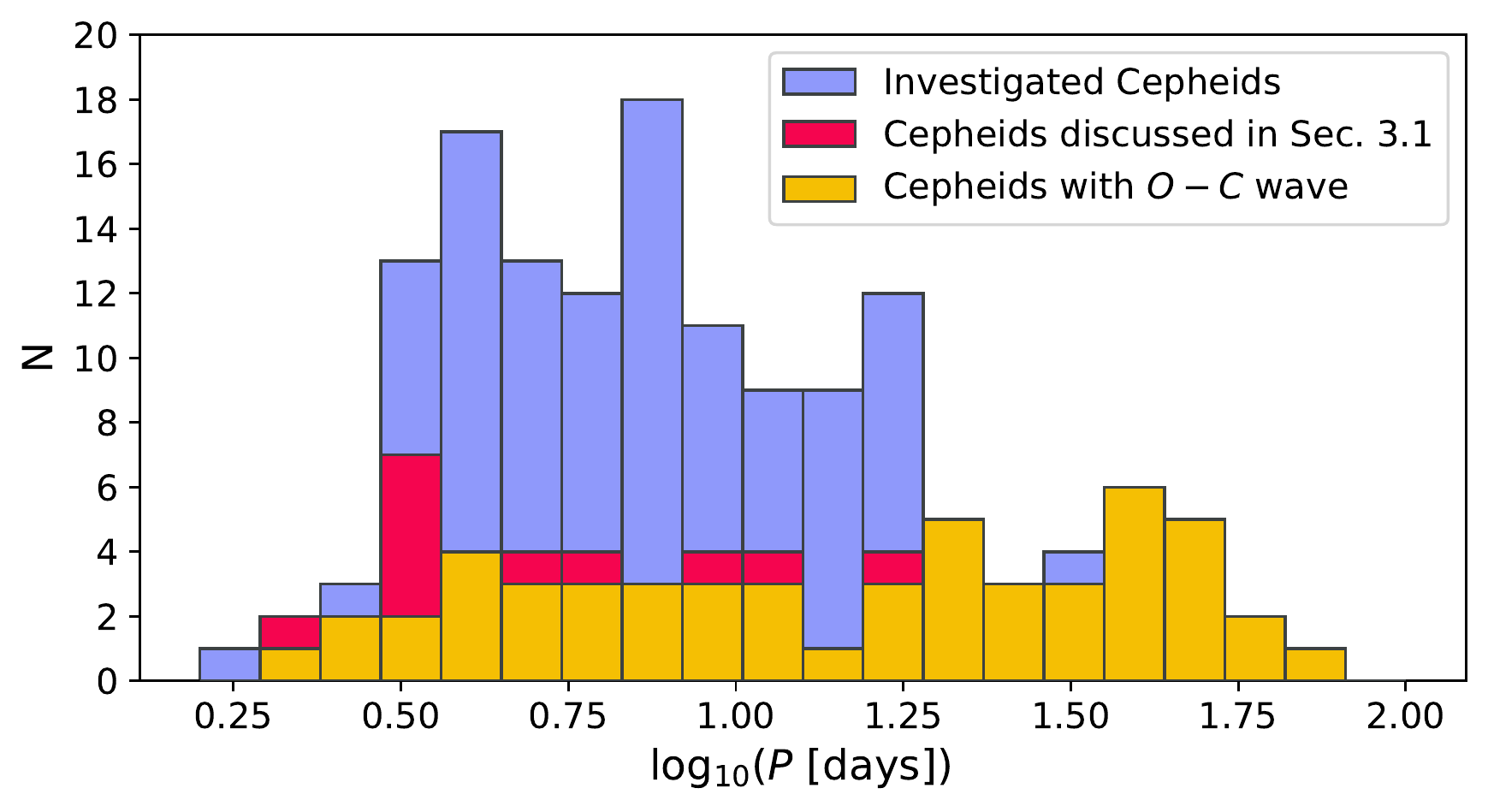}
\caption{Distribution of Cepheids according to their pulsation periods. The blue histogram shows all investigated Cepheids, while the yellow ones show only those that show a wavelike signal in their $O-C$ diagram superimposed the trend corresponding to the evolution. In addition, the red histogram shows the fraction corresponding to the Cepheids termed `peculiar' in the previous sections, while showing wavelike signals in the $O-C$ diagrams.}
\label{fig:flucthist}
\end{figure}

 In Fig.~\ref{fig:flucthist} we plotted the distribution of the investigated Cepheids according to their pulsation periods along with the information about what fraction of Cepheids show fluctuations in their $O-C$ diagram. As seen on the plot, all long period ($\log_{10} P > 1.25$) Cepheids show fluctuations (with the sole exception being EZ~Vel, however the temporal coverage of its photometry did not allow for a detailed fit for the $O-C$ diagram), which was expected based on the behaviour RV~Tauri stars with similar periods. The ratio of stars exhibiting such modulations starts to drop for intermediate period stars and falls to a value of $\sim$25--40\%, which remains approximately the same for shorter period Cepheids too. We suspect that at least a fraction of the stars that were not included in the fluctuating sample also show wavelike modulations of such nature as well, however their overall amplitude was lower than the scatter of the residual $O-C$ diagram, hence they remained below detectability.

It is important to note that the apparently general fact, that all longer period ($\log_{10} P > 1.25$) Cepheids exhibit period fluctuations to some extent, has not been known before, thus it can have important implications for the calibration of the distance ladder via PL relation of Cepheids. Extragalactic Cepheids are generally difficult to observe, as their light curve spans only a few hundred or a thousand days at most, with sparse sampling. Consequently, it is impossible to determine their fluctuation-free, evolution based ``true period'', and the resulting period value can be affected by the cumulative effect of the period fluctuations. Whereas the shorter period ($1.25 < \log_{10} P < 1.48$) population of these Cepheids exhibit only smaller fluctuations and the possible error on the inferred period values remains below a few hundredths of a day, for longer period Cepheids (e.g., for one of the longest period Cepheids in our sample, for S~Vul) such fluctuations could lead to a cumulative effect that alters the inferred period by more than 0.5 day, which would then add a detectable term to the final uncertainty. While such effects might not affect the PL relationship significantly due to the use of a larger sample of stars or the inclusion of shorter period variables (for which the period determination is more accurate) as was done in \cite{Riess2019}, it could severely increase the uncertainty of the results based entirely on long or ultra-long period Cepheids (e.g., \citealt{Bird2009}).

We also measured how the strength and size of these accumulated period fluctuations change for different pulsations periods. For this purpose, we applied the Eddington--Plakidis method (E--P method) \citep{EP} on each Cepheid in our sample that showed signs of fluctuations. According to this method, one must whiten the calculated $O-C$ values for the effect of evolution, then take the absolute values of all delays $u(x) = |a(r+x) - a(r)|$ for each $r$th maximum, $a(r)$, for every possible $x$ cycle separation in the residual dataset. According to \cite{EP}, if the irregular signal is present due to the fluctuation of the phase and period at the same time, then the $\langle u(x)\rangle$ means of all accumulated delays should be related to the random period fluctuation $\epsilon$ by Eq.~\ref{eq:EP}:
\begin{equation}
\label{eq:EP}
\langle u(x)\rangle^2 = 2\alpha^2 + x\epsilon^2
\end{equation}
\noindent where $\alpha$ characterizes the errors in the measured times used for the $O-C$ data points.

\begin{figure} 
\centering
\includegraphics[width = \linewidth]{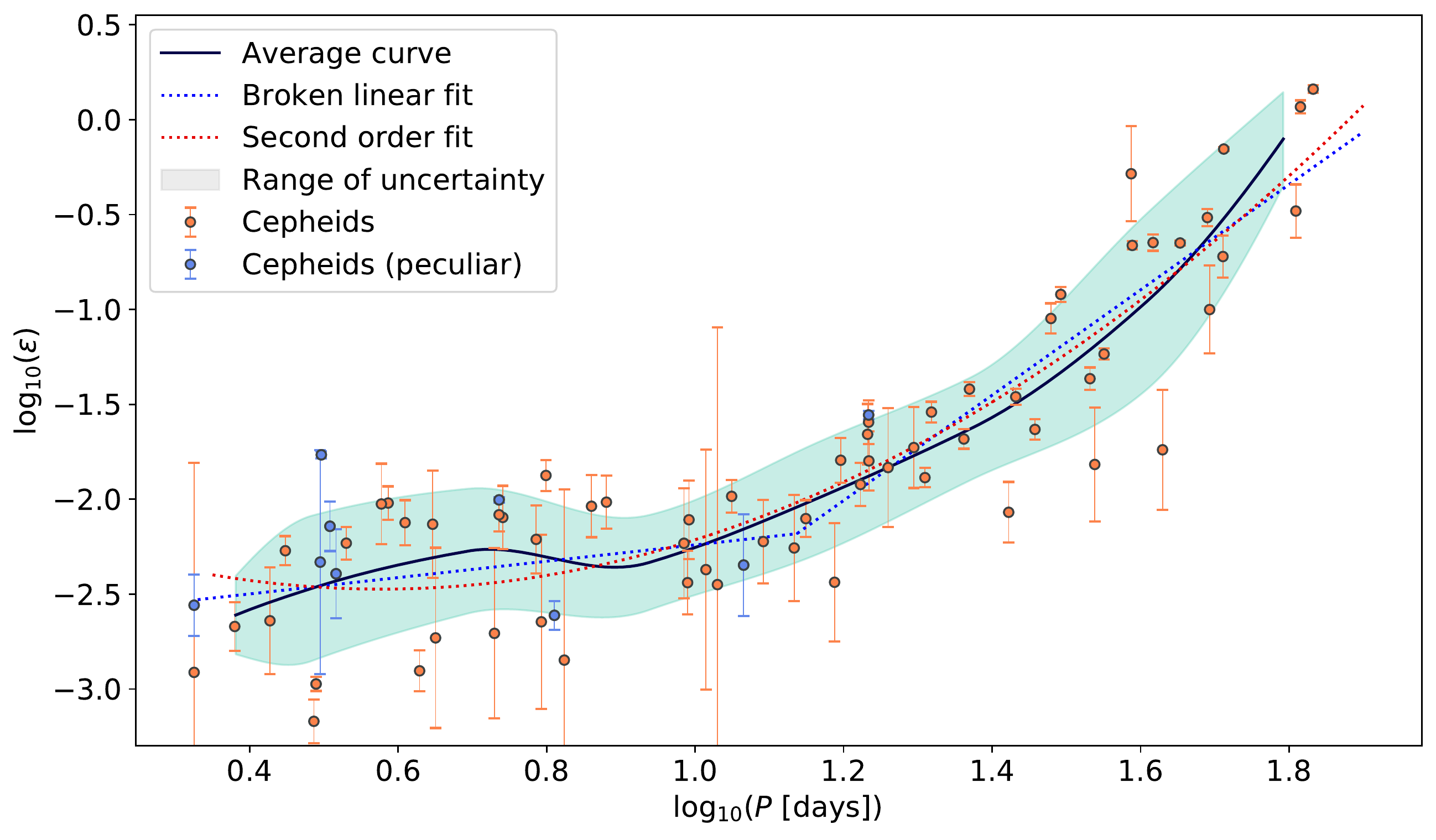}
\caption{Correlation between the Eddington-Plakidis $\epsilon$ parameter calculated for cycle separations $x < 100$ and the pulsation period. The blue and red dashed curves show the two fits described in the text. The black curve shows the empirical average, while the blue region shows the empirical standard deviation around it. For the fitting of the coloured curves, the peculiar Cepheids were not taken into account.}
\label{fig:EPcorr}
\end{figure}

We produced Fig.~\ref{fig:EPcorr} by comparing the resulting $\epsilon$ period fluctuation parameter (obtained by fitting the E--P diagrams for $x < 100$ cycle separations) with the corresponding pulsation period, which shows that there is a significant connection between the size of fluctuations and the pulsation period. 

\begin{figure} 
\centering
\includegraphics[width = \linewidth]{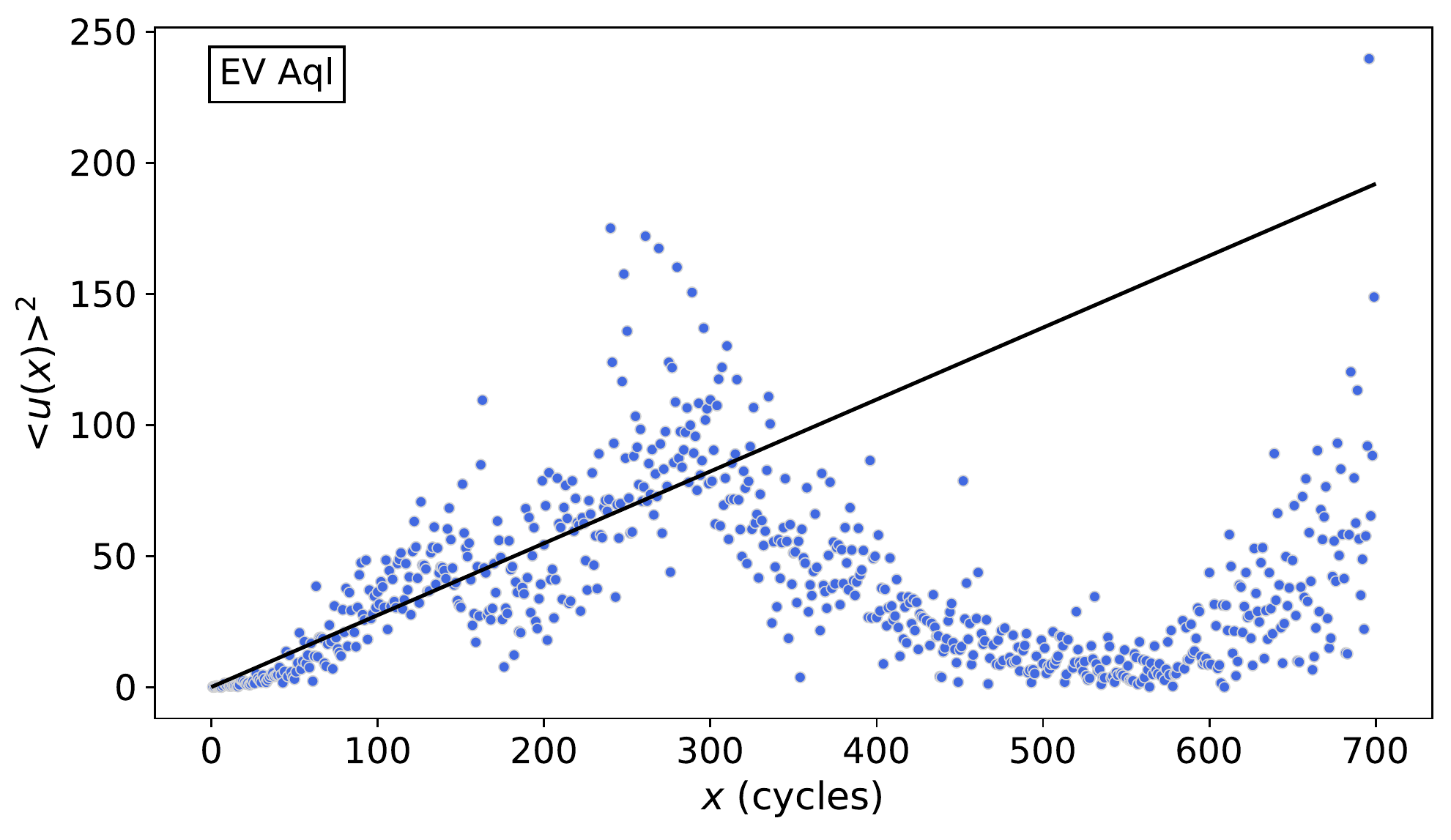}
\caption{The Eddington--Plakidis diagram of EV~Aql, along with the calculated fit for cycle separations $x < 100$. As the diagram shows, multiple choices are possible for this maximal value, which can influence the final result for the period fluctuation parameter.}
\label{fig:EPtest}
\end{figure}

However, the E--P method has two drawbacks which are inherent to its definition, one of which is that one has to manually set what maximal cycle separation is considered for the linear fit described above. There is no statistical background on what one should choose for the maximal cycle separation, and since Cepheids show a very large variation of such E--P diagrams, one has to set this value manually. 
This makes the method infeasible for large datasets, since one either has to tune this value for each star, or pick a general value, which might not be optimal for every sample member. Due to this, the resulting $\epsilon$ fluctuation parameter will also be more uncertain, and its value will be based on a modelling choice. An example for this is also shown in Fig.~\ref{fig:EPtest}: as seen in the figure, multiple ranges could be considered for the fitting of the diagram, which sets a range for the final fluctuation parameter values as well. Additionally, the E-P method also has the disadvantage of not being able to provide an upper limit for the fluctuation parameter, therefore if the $O-C$ diagram is dominated by uncorrelated noise, the resulting E--P diagram will show no structures, unlike in Fig.~\ref{fig:EPtest}. Hence, no linear fitting can be carried out in those scenarios.

To circumvent this issue, we calculated a correlation diagram similar to Fig.~\ref{fig:EPcorr} using the result of the $O-C$ diagram fits instead. For each of the Cepheids where fluctuations were found by the F-test, we simply correlated the measured amplitude of the wavelike modulation with the pulsation period of the star. Unlike in the case of the Eddington-Plakidis method, here we were able to determine a detectability limit even when the amplitude of the modulation could not be measured, i.e., when no fluctuations were found: in this case the detectability limit could still be determined as the maximal value of the Fourier spectrum calculated for the residuals after fitting and subtracting the evolutionary changes.

\begin{figure} 
\centering
\includegraphics[width = \linewidth]{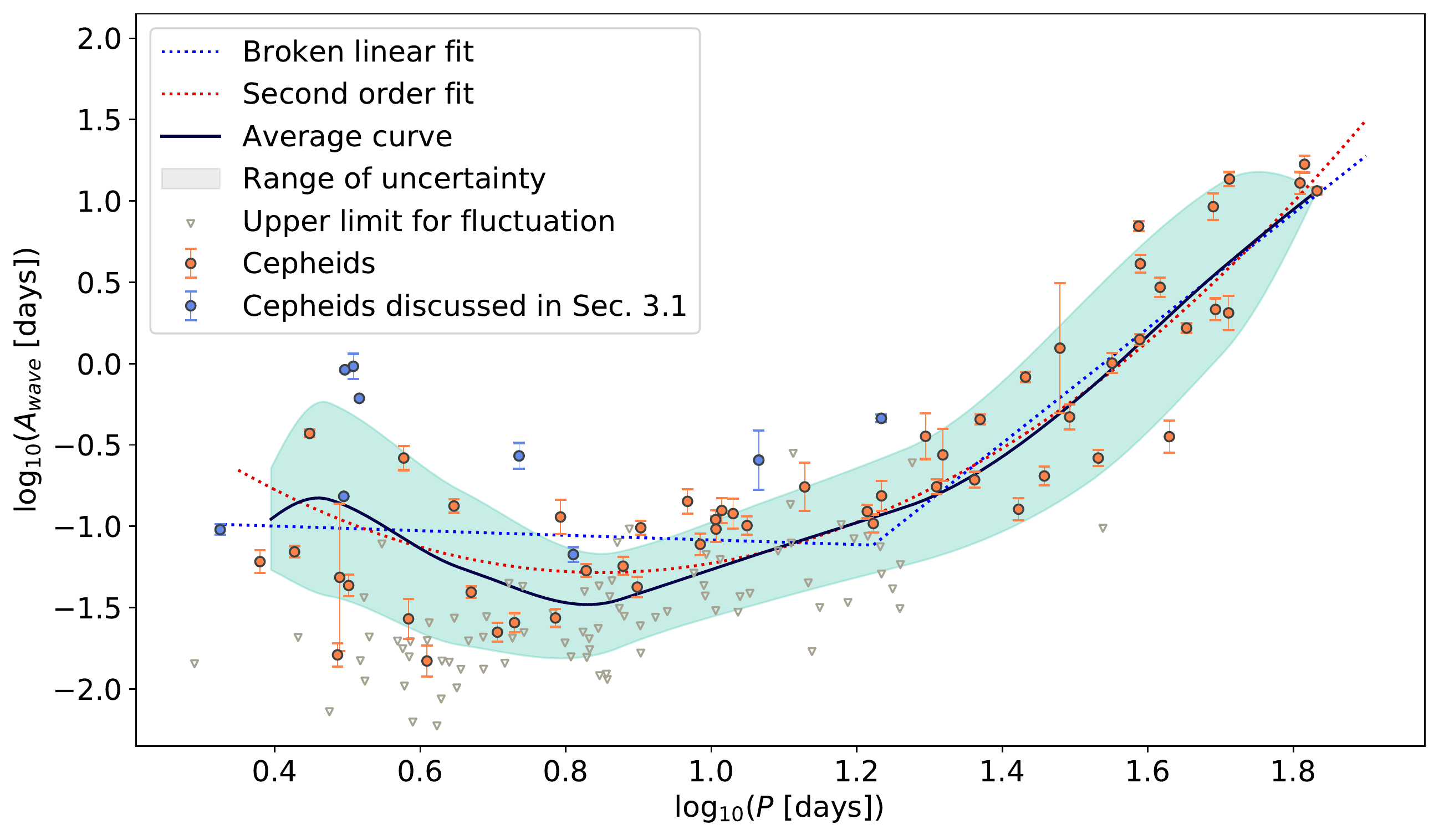}
\caption{Correlation between the measured fluctuation amplitudes and the pulsation periods. The legend is the same as for Fig.~\ref{fig:EPcorr}. The non-detection limits calculated from the Fourier spectrum for individual Cepheids are marked with grey triangles.}
\label{fig:fluctfig}
\end{figure}

Through this procedure we obtained Fig.~\ref{fig:fluctfig}, which shows similar trends as the Eddington--Plakidis method based diagram, except for the shorter period Cepheids. As expected, the detectability limit drawn out by the non-detections increases slightly with period as the light curves of the Cepheids become more and more complex, thus more difficult to fit, but still, a reasonable amount of fluctuating Cepheids were found throughout the investigated period range, which defines the underlying trend reliably. As it is shown by the average curve in Fig.~\ref{fig:fluctfig}, the amplitude of the fluctuations decreases linearly against the logarithm of the period for longer period Cepheids ($\log_{10} P > 1.25$), only to reach a minimum for intermediate period Cepheids around a period of $\log_{10} P \approx 0.8$ and increases again for short period stars. We note that the Cepheids where we found no modulations, and hence are marked as non-detections on the diagram, can significantly influence the position of this minimum. However, the fact that the largest detected fluctuations in the intermediate period range stay well below the maximal ones found for both shorter and longer period Cepheids supports the notion that there is a minimum for these modulations in the period range of bump Cepheids, suggesting a quenching mechanism for these variables. Moreover, Fig.~\ref{fig:fluctfig} also supports the LiTE explanation described above for the modulation found for RX~Aur and Y~Oph, as both stars are outliers for the trend defined by the fluctuations.

We also investigated how easing the F-test based constraints affects the results: if we lower the confidence level from 0.99 to 0.95, the F-test based procedure favours the inclusion of further 23 Cepheids into the fluctuation sample. Most of these stars fall in the period range of bump Cepheids, with the amplitudes of the wavelike signals being close or at the noise level of their $O-C$ diagrams. In case these stars are included, the fraction of Cepheids showing fluctuations jumps to a level of $\sim$50\% instead of 25-40\%, and remains constant thorough the lower end of the pulsation period range, suggesting an even higher frequency for the presence of these modulations. In terms of fluctuation strength, since the fluctuation amplitudes are close to the detection limits assumed for these stars in Fig.~\ref{fig:fluctfig}, which were taken into account for the fitting, neither the actual position of the found fluctuation minimum, nor the general trend changes significantly if we ease the constraints.

As seen in the figures, multiple interpretations are possible: since there are two separate point clouds in the resulting plot, we can either assume that the fluctuation features observed in them are of the same origin, or of different ones. 

By assuming that the fluctuations detected for short and long period Cepheids are of the same origin, the correlation plot can be fitted by a quadratic function. This fit shows that the size of fluctuations decreases from the longer periods towards shorter periods, reaching a minimum at the lower end of the period range of bump Cepheids, then growing back again for shorter period stars. This picture agrees with the presence of a quenching mechanism for bump Cepheids, and it also follows well the average curve.

If one assumes that the two parts of the fluctuation diagram are of different origins (i.e. the fluctuations are of different nature for long and short period stars), then the plot can be fitted by a broken linear curve. One wing of this broken linear curve on the Eddington-Plakidis based diagram (Fig.~\ref{fig:EPcorr}) shows that for long period Cepheids the fluctuations scale linearly with the pulsation period, which agrees with the findings of \cite{Percy1999} for longer period Mira stars, indicating that this fluctuation effect is a similar period fluctuation in both types of variable stars. However, in this case there is no explanation for the short period wing, as this fit would suggest that fluctuations detected for shorter period Cepheids are of different origin than the former one. In the case of the amplitude-based diagram, the broken linear fit shows a similar linear trend for the longer period Cepheids. Based on these fits, the linear scaling present between the fluctuations and the period for long period Cepheids weakens considerably on intermediate and short period ranges, as the level of fluctuations seems not to change as significantly over this range.

The datapoints at short periods ($\log_{10} P < 0.8$) also show a much larger scatter around the model fits compared to other period ranges, with a clear gap in the amplitude distribution (around $\log_{10} A \approx -1.0$, where we find several Cepheids with both significantly larger and smaller fluctuations in this period range). This can indicate that there are more than one type of fluctuations present at short periods, one for normal and one for overtone Cepheids, presumably (with smaller and larger fluctuations, respectively). Their comprehensive modelling would require a more complete sample of shorter period Cepheids, but such an analysis is beyond the scope of this paper. Nevertheless, the supposed models show that either the change in the nature of fluctuation signals occurs at, or the overall strength of the fluctuations have a minimal value close to the period range of the bump Cepheids, both of which suggests that the pulsation of the bump Cepheids is subject to less such random fluctuations and provides an additional proof for the increased stability of the pulsation of these stars.

\section{Summary} 
\label{sec:four}
In this work we presented updated or completely new $O-C$ diagrams for a total of 148 Classical Cepheids. Out of these stars, four Cepheids (RX~Aur, UY~Mon, Y~Oph, and AW~Per) show signs of possible LiTE (only the case of AW~Per and RX~Aur has been known before). Further five Cepheids (XZ~CMa, UX~Car, R~Cru, DT~Cyg, and BN~Pup) exhibit $O-C$ diagrams which either further confirm or raise the question of their binary nature. For the interesting cases of BY~Cas, V532~Cyg, DX~Gem and VZ~CMa we found large amplitude fluctuations influencing the pulsation of the variables, while for IR~Cep we found a similarly large amplitude and complex, yet stable wavelike signal superimposed on the evolutionary changes of the star, which could either be caused by fluctuations or binarity. The proper understanding of these peculiar variations would require further analysis of these stars.

By analysing the obtained set of $O-C$ diagrams showing evolutionary changes, we measured the rates of period change for the individual stars and compared these values to those given by stellar evolution models. The ratio of the number of Cepheids showing positive and negative period changes in our random sample agrees with the ratio given by models that include enhanced mass loss. By plotting the CMD of the obtained sample of Cepheids using the recently published {\em Gaia} EDR3 data, we could show that the rates of period change do not increase as consistently toward the cooler and brighter stars as one would expect based on the models, but the rates of period change vary systematically from the blue edge to the red edge of the instability strip. Considering this, it possibly yields an observational evidence that the rate of period change carries significant information about the physics of these variables.

After fitting and analysing the obtained $O-C$ diagrams, we found that a significant fraction of the Cepheids in our sample exhibit detectable period fluctuations in the form of quasi-periodic waves, similar to those found in the case of other, longer period pulsating stars. By measuring the strength of the fluctuations using the Eddington-Plakidis method and a model fitting based approach, we found that the period fluctuation scales linearly for longer period Cepheids (in a logarithmic sense), which is similar to the behaviour of Mira stars, while for shorter period Cepheids this scaling significantly weakens or disappears. The presence of such fluctuations and their increasing strength for long period Cepheids can have important implications for PL relation based distance determinations, as they could add an additional uncertainty term to the pulsation period, that was previously unaccounted for. Due to the detection limits set by the scatter in the data, we could only detect these fluctuations in about a third of the investigated short and intermediate period Cepheids. We showed that the explanation and analysis of the dependence of fluctuation strength on the periods of short period Cepheids depends largely on whether one or two fluctuation processes are assumed: in the first case the fluctuation strength shows a minimum in the range of bump Cepheids, then it increases back at short periods, while in the latter case we find that the fluctuation strength depends on only weakly or is independent of the period for shorter period stars. Either way, the scatter among the fluctuation strengths increases for short period Cepheids, which suggests a non-negligible contribution from overtone Cepheids that exhibit larger fluctuation signals. Whichever interpretation is correct, the lack of strongly fluctuating bump Cepheids and the fluctuation minimum found by the joint fit in their period range suggests the increased stability of the pulsation of these stars, which is in agreement with the presence of a quenching mechanism.

\section*{Data availability}
All source data used for our analysis are openly available online, except for the measurements carried out at the Piszk\'estet\H{o} Observatory, which are published as online material accompanying the article. The resulting $O-C$ files, from which the conclusions were drawn, are also published in the online supplement and will be accessible in an accompanying Vizier table. The data and the code used for our analysis is available on the GitHub page of the author: \url{https://github.com/Csogeza/O-C_extract}.

\section*{Acknowledgements}
We thank the referee for their useful comments that helped to improve this paper. The research was completed with the extensive use of Python, along with the \texttt{numpy} \citep{numpy}, \texttt{scipy} \citep{scipy} and \texttt{astropy} \citep{astropy} modules, and with the \texttt{R} software \citep{R}. Some of the data used in the article were made available to the community through the Exoplanet Archive on behalf of the KELT project team. This research made use of NASA's Astrophysics Data System (ADS). For the data search and retrieval the authors made extensive use of the Vizier, CDS and Simbad websites of the Strasbourg astronomical Data Center (\url{https://cds.u-strasbg.fr/}). This work has made use of data from the European Space Agency (ESA)
mission {\it Gaia} (\url{https://www.cosmos.esa.int/gaia}), processed by the {\it Gaia} Data Processing and Analysis Consortium (DPAC, \url{https://www.cosmos.esa.int/web/gaia/dpac/consortium}). Funding for the DPAC has been provided by national institutions, in particular the institutions participating in the {\it Gaia} Multilateral Agreement. We acknowledge with thanks the variable star observations from the AAVSO International Database contributed by observers worldwide and used in this research. The research leading to these results has been supported by the Hungarian National Research, Development and Innovation Office (NKFIH) grants K-129249, GINOP-2.3.2-15-2016-00003 and \'Elvonal KKP-137523 (`SeismoLab'), and by the Lend\"ulet LP2018-7/2021 grant of the Hungarian Academy of Sciences. Cs.~K. acknowledges the support provided from the \'UNKP-20-2 and \'UNKP-21-2 New National Excellence Programs of the Ministry of Innovation and Technology from the source of the National Research, Development and Innovation Fund.



\bibliographystyle{mnras}
\bibliography{Citations}



\appendix

\section{Summary table for Cepheids showing evolutionary changes}

In Table~\ref{tab:allceps} we present the period change rates calculated for the Cepheids showing evolutionary period changes along with the data from {\em Gaia} DR3 necessary for the CMD.

\begin{table*}
\renewcommand\thetable{A.1}
\begin{center}
\begin{tabular}{llllllllc}
\hline
Cepheid & P [d] & $\Delta$P|$_{100\textrm{yr}}$ & $g$ & $BP-RP$ & $E(B-V)$ & $\log_{10} \epsilon$ [d] & $\pi$ [mas] & References \\
\hline
U Aql & 7.024 & 2.820e-05 & 9.443 & 1.02 & 0.397 &   & 0.96 & \small{1,2,3,13,53,58,60,64,75,81,A28} \\ 
SZ Aql & 17.141 & 2.267e-04 & 7.986 & 1.755 & 0.614 &   & 0.421 & \small{2,3,12,13,55,58,74,75,81,96,A29} \\ 
TT Aql & 13.755 & -1.223e-05 & 6.945 & 1.744 & 0.4858 &   & 0.888 & \small{1,2,3,12,13,24,53,55,58,60,75,81,97,A29} \\ 
EV Aql & 38.671 & -3.213e-03 & 11.191 & 1.931 & 0.705 & -0.285 & 0.029 & \small{2,3,13,17,49,55,58,81,89,A1} \\ 
FF Aql & 4.471 & 2.582e-05 & 5.143 & 1.148 & 0.221 & -2.731 & 1.81 & \small{2,3,13,53,58,60,74,75,81,85,97,98,A5} \\ 
KL Aql & 6.108 & -3.729e-05 & 10.025 & 1.312 & 0.24 & -2.212 & 0.228 & \small{2,3,13,43,53,55,58,74,79,90,A25} \\ 
V336 Aql & 7.304 & 5.232e-05 & 9.321 & 1.666 & 0.67 &   & 0.47 & \small{2,3,13,53,55,58,74,81,96,A25} \\ 
V493 Aql & 2.988 & 1.407e-05 & 10.572 & 1.7 & 0.777 &   & 0.4 & \small{2,3,13,30,53,55,58,74} \\ 
V496 Aql & 6.807 & 3.075e-05 & 7.351 & 1.438 & 0.442 &   & 0.944 & \small{2,3,13,53,58,74,75,81,A29} \\ 
V916 Aql & 13.443 & 1.045e-04 & 10.044 & 2.186 & 1.089 &   & 0.478 & \small{2,3,13,17,49,55,58,89} \\ 
... & ... & ... & ... & ... & ... & ... & ... & ... \\
\hline
\end{tabular}
\end{center}
\caption{Summary table compiled for the Cepheids showing evolutionary changes. The $\epsilon$ parameter shown in the table corresponds to the period-fluctuation parameter of the Eddington-Plakidis method, which is summarized in Sect.~3.2. The complete table along with the list of photometry references and estimated uncertainties is available in the online supplements of the article.}
\label{tab:allceps}
\end{table*}

\begin{table*}
\renewcommand\thetable{B.1}
\begin{center}
\begin{tabular}{lllllllc}
\hline
Cepheid & $F_{\textrm{H0:lin H1:par}}$ & $F_{\textrm{H0:lin H1:par+wave}}$ & $F_{\textrm{H0:par H1:par+wave}}$ & Amplitude [days] \\
\hline
U Aql & 22.61 & 6.12 & 0.72 &  \\ 
SZ Aql & 60.94 & 17.12 & 1.77 &  \\ 
TT Aql & 6.79 & 10.48 & 11.08 &  \\ 
\bf{EV Aql} & \bf91.78 & \bf315.39 & \bf205.72 & \bf{6.997}\\ 
FF Aql & 63.1 & 24.2 & 7.02 &  \\ 
\bf{KL Aql} & \bf109.7 & \bf70.61 & \bf22.26 & \bf{0.027}\\ 
V336 Aql & 18.55 & 4.79 & 0.39 &  \\ 
V493 Aql & 5.59 & 9.02 & 8.8 &  \\ 
V496 Aql & 9.98 & 3.54 & 1.33 &  \\ 
\bf{V916 Aql} & \bf158.75 & \bf92.54 & \bf23.69 & \bf{0.175}\\ 
... & ... & ... & ... \\
\hline
\end{tabular}
\end{center}
\caption{Summary table of the F-test model comparison conducted for Cepheids showing evolutionary changes: columns 2-3 show the F-statistic and p-value for the case when we compare the residuals of the linear and parabolic fit; columns 4-5 show the results for the linear vs. parabolic plus wave model; columns 6-7 show the results for the parabolic vs. parabolic plus wave model. The last column shows the amplitude of the found wave (if the F-statistic was higher than 10 in the parabolic vs. parabolic plus wave model comparison). When fitting the parabolic plus wave model, we simultaneously optimized for both the parameters of the parabolic change and the wavelike modulation as well. To accept the parabolic plus wave model as a refinement of the simple parabolic fit, we also set the requirement that the found wave should have a period longer than 3000 days (about ten times the folding time used for the $O-C$ diagram calculations) but has to be shorter than five quarters of the timespan covered by observations (to avoid overfitting). Therefore, some of the F-test results were rejected, even though the parabolic plus wave fit seemed more optimal (for example, CH~Cas, XX~Cen or U~Vul). We highlighted the Cepheids that match the criteria with bold fonts. The complete table is available in the online supplement of the article.}
\label{tab:ftest}
\end{table*}

\newpage

\begin{table*}
\renewcommand\thetable{A.2}
\begin{center}
\begin{tabular}{|c|c|c|c|c|c|}
\hline
\# & Reference & \# & Reference & \# & Reference \\
\hline
[1] & AAVSO \citep{aavso}           &        [48] & \cite{Harris1980}            &      [95] & \cite{Szabados1977}        \\

[2] & ASAS \citep{ASAS}             &        [49] & DASCH \citep{DASCH}          &      [96] & \cite{Szabados1980}        \\

[3] & ASAS-SN \citep{asassn1}       &        [50] & \cite{Hellerich1935}         &      [97] & \cite{Szabados1981}         \\

[4] & Abaffy (private comm.)        &        [51] & \cite{Henden1980}            &      [98] & \cite{Szabados1991}         \\

[5] & \cite{Arellano1984}           &        [52] & \cite{Henden1996}            &      [99] & TESS \citep{TESS}           \\

[6] & \cite{Arellano1998}           &        [53] & Hipparcos \citep{hipp}       &      [100] & Tabur (private comm.)       \\

[7] & \cite{Arp1959}                &        [54] & \cite{Hoffmeister1960}       &      [101] & \cite{Bahner1971}           \\

[8] & \cite{Asarnova1957}           &        [55] & IOMC \citep{IOMC}            &      [102] & \cite{Usenko1992}           \\

[9] & \cite{Babel1989}              &        [56] & \cite{Irwin1961}             &      [103] & \cite{Wachmann1976}         \\

[10] & \cite{Bahner1962}            &        [57] & Kepler K2 \citep{K2}         &      [104] & \cite{Walraven1958}        \\

[11] & \cite{Barnes1987}            &        [58] & KWS \citep{kws}              &      [105] & \cite{Walraven1964}        \\

[12] & \cite{Barnes1997}            &        [59] & KELT \citep{Kelt}            &      [106] & \cite{Walter1943}          \\

[13] & \cite{Berdnikov2008}         &        [60] & \cite{Kiss1998}              &      [107] & \cite{Weaver1960}          \\

[14] & \cite{Berdnikov2009}         &        [61] & \cite{Koukarkina1954}        &      [108] & Williams (private comm.)   \\

[15] & \cite{Berdnikov2011}         &        [62] & \cite{Kovacs1979}            &      [A1] & \cite{Berdnikov1994}        \\

[16] & \cite{Berdnikov2014}         &        [63] & \cite{Kox1935}               &      [A2] & \cite{Berdnikov1999}        \\

[17] & \cite{Berdnikov2015}         &        [64] & \cite{Krebs1936}             &      [A3] & \cite{Berdnikov2000}        \\

[18] & \cite{Berdnikov2019}         &        [65] & \cite{Kurochkin1954}         &      [A4] & \cite{Berdnikov2004}        \\

[19] & \cite{Bersier2002}           &        [66] & \cite{Kwee1967}              &      [A5] & \cite{Berdnikov2014}        \\

[20] & Bochum \citep{Bochum}         &        [67] & \cite{Landolt1971}           &      [A6] & \cite{Berdnikov2019b}        \\

[21] & \cite{Buchancowa1972}        &        [68] & \cite{Laur2017}              &      [A7] & \cite{Erleksova1978}        \\

[22] & \cite{Burnashev2009}         &        [69] & \cite{Madore1975}            &      [A8] & \cite{Erleksova1982}        \\

[23] & \cite{Connolly1983}          &        [70] & \cite{Malik1965}             &      [A9] & \cite{Fernie1990}           \\

[24] & \cite{Coulson1985}           &        [71] & \cite{Mauder-Schoeffel1968}  &      [A10] & \cite{Hacke1989}            \\

[25] & \cite{Cousins1968}           &        [72] & \cite{Millis1969}            &      [A11] & \cite{Hacke1990}           \\

[26] & \cite{Cousins1971}           &        [73] & \cite{Mitchell1961}          &      [A12] & \cite{Heiser1996}          \\

[27] & \cite{Dean1977}              &        [74] & \cite{Mitchell1964}          &      [A13] & \cite{Kiehl1977}           \\

[28] & \cite{Dean1981}              &        [75] & \cite{Moffett1984}           &      [A14] & \cite{Klawitter1971}       \\

[29] & \cite{Detre1935}             &        [76] & NSVS \citep{NSVS}            &      [A15] & \cite{MahmoudSzabados1980} \\

[30] & \cite{Diethelm1982}          &        [77] & \cite{OGLE}                  &      [A16] & \cite{Meyer2006}           \\

[31] & \cite{Eggen1951}             &        [78] & \cite{Oja2011}               &      [A17] & \cite{Miller-Wachmann1973} \\

[32] & \cite{Eggen1957}             &        [79] & \cite{Oosterhoff1960}        &      [A18] & \cite{Nijland1935}         \\

[33] & \cite{Eggen1983}             &        [80] & \cite{Parkhurst1908}         &      [A19] & \cite{Oosterhoff1935}      \\

[34] & \cite{Eggen1985}             &        [81] & \cite{Pel1976}               &      [A20] & \cite{Oosterhoff1943}      \\

[35] & \cite{Feinstein1969}         &        [82] & \cite{Pingsdorf1935}         &      [A21] & \cite{Parenago1956}        \\

[36] & \cite{Fernie1965}            &        [83] & Present paper                &      [A22] & \cite{Romano1958}          \\

[37] & \cite{Fernie1970}            &        [84] & \cite{Reed1968}              &      [A23] & \cite{Strohmeier1968}      \\

[38] & \cite{Fernie1979}            &        [85] & SMEI \citep{SMEI}            &      [A24] & \cite{Szabados1977}        \\

[39] & \cite{Fernie1981}            &        [86] & SWASP \citep{SWASP}          &      [A25] & \cite{Szabados1980}        \\

[40] & \cite{Fernie1995}            &        [87] & \cite{Schmidt1993}           &      [A26] & \cite{Szabados1981}        \\

[41] & \cite{Filatov1957}           &        [88] & \cite{Schmidt1995}           &      [A27] & \cite{Szabados1988}        \\

[42] & \cite{Floria1953}            &        [89] & \cite{Schmidt2004}           &      [A28] & \cite{Szabados1989}        \\

[43] & Gaia DR2 \citep{Gaia2}       &        [90] & \cite{Schmidt2005}           &      [A29] & \cite{Szabados1991}        \\

[44] & \cite{Gaposchkin1958}        &        [91] & \cite{Shobbrook1992}         &      [A30] & \cite{Szabados2013}        \\

[45] & \cite{Gieren1981}            &        [92] & \cite{Stobie1970}            &      [A31] & \cite{Vinko1991}           \\

[46] & \cite{Gieren1985}            &        [93] & \cite{Stobie1979}            &      [A32] &  \cite{Walker1991}         \\

[47] & \cite{Grayzeck1978}          &        [94] & \cite{Szabados1976}          &            &                            \\
\hline
\end{tabular}
\end{center}
\caption{Complete list of references used for the calculation of the $O-C$ diagrams along with the abbreviations.}
\end{table*}

\bsp	
\label{lastpage}
\end{document}